\documentstyle[12pt,epsfig,amsfonts]{article}
\newcommand{\beq}{\begin{eqnarray}}
\newcommand{\eeq}{\end{eqnarray}}
\newcommand{\eqn}{\begin{equation}}
\newcommand{\een}{\end{equation}}
\begin{document}
\title{Semi-classical mass of quantum $k$-component topological kinks}
\author{A. Alonso Izquierdo$^{(a)}$, W. Garc\'{\i}a Fuertes$^{(b)}$,
M.A. Gonz\'alez Le\'on$^{(a)}$ \\ and J. Mateos Guilarte$^{(c)}$
\\ {\normalsize {\it $^{(a)}$ Departamento de Matem\'atica
Aplicada}, {\it Universidad de Salamanca, SPAIN}}\\{\normalsize
{\it $^{(b)}$ Departamento de F\'{\i}sica}, {\it Universidad de
Oviedo, SPAIN}}\\ {\normalsize {\it $^{(c)}$ Departamento de
F\'{\i}sica}, {\it Universidad de Salamanca, SPAIN}}}

\date{}
\maketitle
\begin{abstract}
We use the generalized zeta function regularization method to
compute the one-loop quantum correction to the masses of the TK1
and TK2 kinks in a deformation of the $O(N)$ linear sigma model on
the real line.
\end{abstract}
\clearpage
\section{Introduction}

In this paper we shall apply the generalized zeta function
regularization method to unveil the semi-classical behaviour of
the quantum topological kinks arising as BPS states in
multi-component scalar (1+1)-dimensional field theory. The
interest of this investigation lies in the fact that these systems
live at the heart of the low energy regime of string/M theory.
Effective theories such as ${\cal N}$=1 SUSY QCD and/or the
Wess-Zumino model \cite{Shifman} encompass $N$-component scalar
fields and have a discrete set of vacuum states. From a
one-dimensional perspective one can foresee the existence of
$k$-component topological kinks, $k\leq N$, which are the seeds of
the BPS domain walls. These extended states play such a prominent
r$\hat{\rm o}$le in the three-dimensional world that analysis of
the quantum behaviour of $k$-component topological kinks becomes
an important issue.

Here, we shall focus on a one-parametric family of deformations of
the $O(N)$ linear sigma model, in (1+1)-dimensional space-time.
The model to be addressed forms the bosonic sector of a
super-symmetric theory with $N$ real super-fields and is of the
general type of the Wess-Zumino model, like those studied in
\cite{Shifman} and \cite{Losev}. An important feature common to
all these models is the non-existence of continuous symmetries in
internal space and they can therefore be included in the class of
systems considered in Reference \cite{Graham}. The authors of
these works use continuous phase-shift methods to calculate the
one-loop correction to the kink energy in full super-symmetric
theories.

Our approach differs in two ways: 1) we restrict ourselves to the
bosonic sector and leave the fermionic fluctuations for future
research; 2) the generalized zeta function regularization
procedure is applied to deal with the infinite quantities arising
in the quantization prescription. This method has been used
previously in the description of quantum corrections to kink
masses for theories with a single real scalar field, see
\cite{Bordag}, but here we shall follow the more elaborated
procedure developed in \cite{AGM} and \cite{Bor} for one-component
systems. Nevertheless, our method is particularly suited to models
with several real fields because in general, in theories of this
type, the spectrum of the second order fluctuation operator around
kink solutions is only partially (asymptotically) known. Since
this operator is a Schr\"odinger operator acting on functions
belonging to the Hilbert space $L^2({\Bbb R}) \otimes {\Bbb C}^N$,
one can write an associated heat equation. From the asymptotic
expansion to the heat function one obtains enough information
about the generalized zeta function of the second fluctuation
operator \cite{Gilkey} to provide a good approximation to the
one-loop correction to the kink mass. Asymptotic methods to
compute the mass of quantum solitons were first used in Reference
\cite{Dunne}.

Within a given range of the deformation parameter, the modified
$O(2)$ linear sigma model that we shall consider is the celebrated
Montonen-Sarker- Trullinger- Bishop ( MSTB ) model
\cite{Montonen}. Over the years, several kinds of kink solitary
waves with very noticeable properties have been discovered in this
system , \cite{Rajar}. The kink moduli space of the $N\geq 3$
generalizations of the MSTB model has been described in
\cite{AGM4}, whereas more recently the stability of the different
kinds of kink was established in \cite{AGM2}. Therefore,
computation of the semi-classical mass of the stable topological
kinks with one non-null component, TK1 , or two non-null
components, TK2, is compelling. To achieve this task, which is the
main goal of this paper, the zeta function regularization method
is specially appropriate , because of the impossibility of solving
the spectral problem of the second order fluctuation operator.

The organization of the paper is as follows: In Section \S.2 the
general semi-classical formula for the mass of quantum solitons,
the zeta function regularization procedure, and the zero point
energy and mass renormalization prescriptions are explained. In
Section \S.3 we compute the one-loop quantum corrections to the
masses of topological kinks with one and two non-null components
in the MSTB model. In Section \S.4 similar formulas are given in
the deformation of the linear $O(N)$ sigma model, with $N\geq 3$,
that generalizes the MSTB model. Section \S.5 offers some comments
on possible extensions of our results. Finally, in Appendix A the
first coefficients of the matrix heat kernel expansion are
written, whereas, in Appendix B, it is shown that only the stable
two-component topological kinks saturate the topological bound.

\section{Semi-classical mass formula for k-component quantum topological kinks}

$N$-component scalar fields are maps from the ${\Bbb R}^{1,1}$
Minkowski space-time to the ${\Bbb R}^N$ \lq\lq internal" space:
\[
\vec{\psi}(y^\mu)=\sum_{a=1}^N\psi_a(y^\mu)\vec{e}_a:{\Bbb
R}^{1,1}\longrightarrow {\Bbb R}^N  .
\]
Here, $y^\mu$, $\mu=0,1$, are coordinates in ${\Bbb R}^{1,1}$;
$\partial_\mu=\frac{\partial}{\partial y^\mu}$ is a basis in $T
{\Bbb R}^{1,1}$, and $\vec{e}_a, a=1,2,\cdots ,N$, are
ortho-normal vectors in ${\Bbb R}^N$, $\vec{e}_a \cdot
\vec{e}_b=\delta_{ab}$. We shall consider (1+1)-dimensional field
theories whose classical dynamics is governed by the action
\[
S=\int d^2 y \left\{ \frac{1}{2} \, \partial_\mu \vec{\psi} \,
\partial^\mu \vec{\psi}-U(\vec{\psi}) \right\}.
\]
We choose the metric tensor in $T^2({\Bbb R}^{1,1})$ as $g={\rm
diag}\,(1,-1)$ and the Einstein convention will be used throughout
the paper only for the indices in ${\Bbb R }^{1,1}$. The system of
units -identical to that chosen in \cite{AGM}- is such that only
the speed of light is set to $c=1$.

The classical configuration space ${\cal C}$  is formed by the
static configurations $\vec{\psi}(y)$ - we denote the spatial
coordinate as $y^1=y$ - for which the energy functional
\[
E(\vec{\psi})=\int dy \left\{ \frac{1}{2} \sum_{a=1}^N \frac{d
\psi_a}{d y} \frac{d \psi_a}{d y}+U(\vec{\psi}) \right\}
\]
is finite: ${\cal C}=\{\vec{\psi}(y) \,/\, E(\vec{\psi}) <
+\infty\}$. In the Schr\"odinger picture, the quantum evolution is
ruled by the Schr\"odinger functional equation
\[
i \hbar \frac{\partial}{\partial t} \Psi[\vec{\psi}(y),t=y^0]=
{\Bbb H} \Psi[\vec{\psi}(y),t=y^0].
\]
If $\vec{\psi}_V$ is a constant minimum of $U$, the masses of the
fundamental quanta are: $v_a^2\delta_{ab}= \frac{\delta^2
U}{\delta \psi_a \delta \psi_b} |_{\vec{\psi}_V}$. The quantum
Hamiltonian operator
\[
{\Bbb H}=\int dy \left\{-\frac{\hbar^2}{2} \sum_{a=1}^N v_a^2
\frac{\delta}{\delta \psi_a(y)} \frac{\delta}{\delta \psi_a(y)}
\right\} +E[\vec{\psi}(y)]
\]
acts on wave functionals $\Psi[\vec{\psi}(y),t]$ that belong to
$L^2({\cal C})$.

A straightforward generalization of the arguments and definitions
of the Section \S .2 of \cite{AGM} - following the classical
papers \cite{Dashen}, \cite{Kor} - shows that the kink ground
state energy at the semi-classical limit is:
\begin{equation}
{\Bbb E}_0^{\rm K}=E[{\vec\psi}_{\rm K}]+\frac{\hbar}{2} \, {\rm
Tr}\, (P{\cal K})^{\frac{1}{2}}+o(\hbar^2)=E[{\vec\psi}_{\rm
K}]+\frac{\hbar}{2} \, \zeta_{\rm P{\cal
K}}(-\textstyle\frac{1}{2})+o(\hbar^2) \quad .\label{eq:dosdos}
\end{equation}
Here ${\cal K}$ is the second variation operator, with spectral
equation ${\cal K}\vec{\xi}_n(x)=\omega_n^2\vec{\xi}_n(x)$, and
\[
\zeta_{\rm P{\cal K}}(s)=\sum_{\omega_n^2>0}
\frac{1}{(\omega_n^2)^s}
\]
is the associated generalized zeta function; $P$ is the projector
to the strictly positive part of ${\rm Spec}\hspace{0.1cm}{\cal
K}$. In this case, however, ${\cal K}$ is a $N\times N$ \lq\lq
matrix" differential operator with \lq\lq matrix elements"
\[
{\cal K}_{ab}=-\frac{\partial^2}{\partial y^2}\,
\delta_{ab}+\left. \frac{\delta^2 U}{\delta \psi_a \delta \psi_b}
\right|_{\vec{\psi}_{\rm K}},\ \ \ \     a,b=1,2,\cdots,N\, .
\]

\subsection{Generalized zeta function regularization method}
We shall regularize $ \zeta_{P {\cal K}} (-\frac{1}{2})$ by
defining the analogous quantity $ \zeta_{P {\cal K}} (s)$ at some
point in the complex $s$-plane where $\zeta_{P {\cal K}}(s)$ does
not have a pole. $\zeta_{P {\cal K}}(s)$ is a meromorphic function
of $s$ such that its residues and poles can be derived through
heat kernel methods, see \cite{Gilkey}. If $K_{\rm {\cal
K}}(y,z;\beta)$ is the kernel of the $N\times N$ \lq\lq matrix"
heat equation associated to ${\cal K}$,
\begin{equation}
\left( \frac{\partial}{\partial \beta}{\bf 1}_N +{\cal K} \right)
K_{\rm {\cal K}}(y,z;\beta)=0 \hspace{0.4cm}, \hspace{0.4cm}
K_{\rm {\cal K}}(y,z;0)={\bf 1}_N\delta(y-z)\, , \label{eq:eqcalor}
\end{equation}
the Mellin transformation tells us that
\[
\zeta_{P {\cal K}}(s)=\frac{1}{\Gamma(s)} \int_0^\infty d\beta \,
\beta^{s-1} \, {\rm Tr} \, e^{-\beta \, P {\cal K}}\quad ,
\]
where
\[
h_{P{\cal K}}[\beta]={\rm Tr}\, e^{-\beta\, P {\cal K}}={\rm Tr}\,
e^{-\beta {\cal K}}-j=-j+ \int_{-\infty}^{\infty} dy \,K_{\rm {\cal K}}(y,y;\beta)
\]
is the heat function $h_{P {\cal K}}[\beta]$,  if ${\cal K}$ is
positive semi-definite and ${\rm dim Ker}{\cal K}=j$. The \lq\lq
regularized" kink energy is at the semi-classical limit:
\begin{equation}
{\Bbb E}_0^{\rm K}(s)=E[{\phi}_{\rm K}]+\frac{\hbar}{2} \mu^{2s+1}
\zeta_{P{\cal K}}(s)+ o(\hbar^2) \label{eq:ener}
\end{equation}
where $\mu$ is a unit of ${\rm length}^{-1}$ introduced to render
the terms in (\ref{eq:ener}) homogeneous from a dimensional point
of view. The infiniteness of the bare quantum energy shows itself
here in the pole that the zeta function develops for
$s=-\frac{1}{2}$.

The renormalization of ${\Bbb E}_0^K(s)$ will be performed in the
same three steps as in \cite{AGM}:

A. The quantum fluctuations around the vacuum are governed by the
Schr\"odinger operator:
\[
{\cal V}_{ab}=-\frac{d^2}{dy^2} \, \delta_{ab}+\frac{\delta^2
U}{\delta \psi_a \delta \psi_b}|_{\vec{\psi}_V}
\]
where $\vec{\psi}_V$ is a constant minimum of $U$ and ${\cal
U}_{ab}(\vec{\psi}_V)= \frac{\delta^2 U}{\delta \psi_a \delta
\psi_b} |_{\vec{\psi}_V}=v_a^2 \delta_{ab}$ is the matrix of
second variational derivatives of $U$ at $\vec{\psi}_V$. The
kernel of the heat equation
\[
\left( \frac{\partial}{\partial \beta}+{\cal V} \right)K_{\cal
V}(y,z;\beta)=0 \hspace{0.4cm}, \hspace{0.4cm} K_{\cal
V}(y,z;0)={\bf 1}_N \, \delta(y-z)
\]
provides the heat function $h_{\cal V}(\beta)$,
\[
h_{\cal V}(\beta)={\rm Tr}\, e^{-\beta {\cal V}}=\sum_{a=1}^N \int_{-\infty}^\infty
dy K_{\cal V}(y,y;\beta)
\]
and, through the Mellin transformation, we obtain
\[
\zeta_{\cal V}(s)=\frac{1}{\Gamma(s)} \int_0^\infty d\beta \,
\beta^{s-1}\, {\rm Tr} \, e^{-\beta {\cal V}}.
\]
The regularized kink energy measured with respect to the
regularized vacuum energy is thus
\begin{eqnarray*}
{\Bbb E}^K(s)&=&E[\vec{\psi}_{\rm
K}]+\Delta_1\varepsilon^K(s)+o(\hbar^2)\\&=&E[\vec{\psi}_{\rm
K}]+\frac{\hbar}{2} \mu^{2s+1}\left[\zeta_{P{\cal
K}}(s)-\zeta_{\cal V}(s) \right]+o(\hbar^2).
\end{eqnarray*}

B. If we now pass to the physical limit ${\Bbb E
}^K=\lim_{s\rightarrow -\frac{1}{2}} {\Bbb E}^K(s)$, we still
obtain an infinite result. The reason for this is that the
physical parameters of the theory have not been renormalized. It
is well known that in (1+1)-dimensional scalar field theory normal
ordering takes care of all renormalizations in the system: the
only ultraviolet divergences that occur in perturbation theory
come from graphs that contain a closed loop consisting of a single
internal line, \cite{Coleman} . From Wick's theorem, adapted to
contractions of two fields at the same point in space-time, we see
that normal ordering adds to the Hamiltonian the mass
renormalization counter-terms

\[
H(\delta m^2)=-\frac{\hbar}{2} \int dy \sum_{a=1}^N \delta
m_{aa}^2 :\frac{\delta^2 U}{\delta \psi_a \delta
\psi_a}:+o(\hbar^2)
\]
up to one-loop order. To regularize
\[
\delta m_{aa}^2= \int \frac{dk}{4\pi}
\frac{\delta_{aa}}{\sqrt{k^2+{\cal U}_{aa}(\vec{\psi}_V)}}\quad ,
\]
we first put the system in a 1D box of length $L$ so that
$\delta m_{aa}^2=\frac{1}{2L}\zeta_{{\cal V}_{aa}}(\frac{1}{2})$, if the
constant eigen-function of ${\cal V}_{aa}$ is not included in
$\zeta_{{\cal V}_{aa}}$. Then, we again use the zeta function
regularization method and define: $\delta
m_{aa}^2(s)=-\frac{1}{L}\frac{\Gamma (s+1)}{\Gamma (s)}\mu^{2s+1} \zeta_{{\cal
V}_{aa}}(s+1)$. Note that $\delta m_{aa}^2=\lim_{s\to -\frac{1}{2}}\delta
m_{aa}^2(s)$. The criterion behind this regularization prescription is
the vanishing tadpole condition, which is shown in Appendix B of
Reference \cite{Bor} to be equivalent to the heat kernel
subtraction scheme.

The one-loop
correction to the kink energy due to $H(\delta m^2(s))$ is thus
\begin{eqnarray}
\Delta_2 \varepsilon^K(s)&=&\left<\vec{\psi}_K \right| H(\delta
m^2(s)) \left| \vec{\psi}_K \right>-\left<\vec{\psi}_V \right|
H(\delta m^2(s)) \left| \vec{\psi}_V \right> = \nonumber\\ &=&
\lim_{L\rightarrow\infty}\frac{\hbar}{2L} \mu^{2s+1}\frac{\Gamma(s+1)}{\Gamma(s)}
\int_{-\frac{L}{2}}^{\frac{L}{2}} dy \sum_{a=1}^N \zeta_{{\cal
V}_{aa}}(s+1) \left[ \left. \frac{\delta^2 U}{\delta \psi^a \delta
\psi_a} \right|_{\vec{\psi}_K}-  \left. \frac{\delta^2 U}{\delta
\psi^a \delta \psi_a} \right|_{\vec{\psi}_V} \right]\nonumber
\\&=&-\lim_{L\rightarrow\infty}\frac{\hbar}{2L} \mu^{2s+1}\frac{\Gamma(s+1)}{\Gamma(s)}\sum_{a=1}^N
\zeta_{{\cal V}_{aa}}(s+1)\int_{-\infty}^\infty dy  V_{aa}(y)
\end{eqnarray}
because the expectation values of normal ordered operators in
coherent states are the corresponding c-number-valued functions.

 C. Finally, the renormalized kink energy is :

\begin{equation}
{\Bbb E}_R^K=E[\vec{\psi}_K]+ \lim_{s\rightarrow -\frac{1}{2}}
\left[ \Delta_1 \varepsilon^K(s)+ \Delta_2\varepsilon^K(s)
\right]+o(\hbar^2)\quad .\label{eq:w1}
\end{equation}

\subsection{Asymptotic approximation to semi-classical kink masses}

In general it is very difficult to compute $h_{\rm P{\cal
K}}[\beta]$ exactly. In such a case, we shall make use of the
asymptotic expansion of $h_{\rm P{\cal K}}[\beta]$, which is well
defined if $0<\beta<1$. In order to use the asymptotic expansion
of the generalized zeta function of the ${\cal K}$ operator to
compute the semi-classical expansion of the corresponding quantum
kink mass, it is convenient to use non-dimensional variables. We
define non-dimensional space-time coordinates $x^\mu=m_dy^\mu$ and
field amplitudes $\vec{\phi}(x^\mu)=c_d\vec{\psi}(y^\mu)$, where
$m_d$ and $c_d$ are constants with dimensions $[m_d]=L^{-1}$ and
$[c_d]=M^{-\frac{1}{2}}L^{-\frac{1}{2}}$. Also, we write
$\bar{U}(\vec{\phi})=\frac{c_d^2}{m_d^2}U(\vec{\psi})$.

The action and the energy can now be written in terms of their
non-dimensional counterparts:
\begin{eqnarray*}
S[\vec{\psi}]&=&\frac{1}{c_d^2}\int
d^2x\left(\frac{1}{2}\frac{\partial \vec{\phi}}{\partial
x_\mu}\frac{\partial \vec{\phi}}{\partial
x^\mu}-\bar{U}(\vec{\phi})\right)=\frac{1}{c_d^2}\bar{S}[\vec{\phi}]\\
E[\vec{\psi}]&=&\frac{m_d}{c_d^2}\int dx\left(\frac{1}{2}\frac{d
\vec{\phi}}{dx}\frac{d
\vec{\phi}}{dx}+\bar{U}(\vec{\phi})\right)=\frac{m_d}{c_d^2}\bar{E}[\vec{\phi}].
\end{eqnarray*}
The important point is that the Hessians at the vacuum and kink
configurations can always be written as the $N \times N$ matrix
differential operators ${\displaystyle \bar{\cal V}
=\frac{1}{m_d^2}{\cal V}}$ and ${\displaystyle\bar{\cal K}
=\frac{1}{m_d^2}{\cal K}}$, given by
\begin{eqnarray*}
\bar{\cal V}_{ab}&=&-\frac{d^2}{d x^2}\delta_{ab}+\bar{v}_a^2\delta_{ab}\\
\bar{\cal K}_{ab}&=&-\frac{d^2}{d x^2}\delta_{ab}+\bar{v}_a^2\delta_{ab}-\bar{V}_{ab}(x)
\end{eqnarray*}
where
${\displaystyle \frac{\delta^2\bar{U}}{\delta\phi_a^2}|_{\vec{\phi}_V}=\bar{v}_a^2}$
and
${\displaystyle\frac{\delta^2\bar{U}}{\delta\phi_a\delta\phi_b}|_{\vec{\phi}_K}=\bar{v}_a^2\delta_{ab}
-\bar{V}_{ab}(x)}$. Therefore,
\[
\zeta_{\cal V}(s)=\frac{1}{m_d^{2s}}\zeta_{\bar{\cal
V}}(s)\hspace{0.5cm},\hspace{0.5cm}\zeta_{\cal
K}(s)=\frac{1}{m_d^{2s}}\zeta_{\bar{\cal K}}(s).
\]
The asymptotic expansion of $N\times N$ matrix heat
kernels is given in Reference \cite{Avra}. Nevertheless, we shall
sketch this procedure in order to adapt it to our computational
needs. We thus write the heat kernel in the form
\[
K_{\bar{\cal K}}(x,x';\beta)=A(x,x';\beta)K_{\bar{\cal
V}}(x,x';\beta)
\]
where
\[
[K_{\bar{\cal V}}]_{ab}(x,x';\beta)=\frac{1}{\sqrt{4\pi \beta}}\exp\left[-\frac{(x-x')^2}{4\beta}\right] \exp(-\beta\bar{v}_a^2)\delta_{ab}
\]
is the solution of the heat kernel equation for the $\bar{\cal V}$
operator with the initial condition $[K_{\bar{\cal
V}}]_{ab}(x,x';0)= \delta_{ab}\delta(x-x')$. The matrix elements
of the $N\times N$ matrix-valued function $A(x,x';\beta)$ satisfy
the system of $N^2$ coupled PDE:
\begin{equation}
\sum_{c=1}^N\sum_{d=1}^N\left\{\left(\frac{\partial}{\partial\beta}
-\frac{\partial^2}{\partial x^2}+\bar{v}_a^2
\right)\delta_{ac}-\bar{V}_{ac}(x)\right\} [A]_{cd}(x,x')
[K_{\bar{\cal V}}]_{db}(x,x';\beta)=0 \quad , \label{mateq}
\end{equation}
whereas $A(x,x';0)={\bf 1}_N$ is the $N\times N$ identity matrix.

For $\beta<1$, we solve (\ref{mateq}) by means of an asymptotic
(high-temperature) expansion: $A(x,x';\beta)=\sum_{n=0}^\infty
a_n(x,x') \beta^n$. In this regime the heat function is given by:
\begin{eqnarray*}
{\rm Tr}\, e^{-\beta \bar{\cal K}} &=&
\sum_{a=1}^N\int_{-\infty}^{\infty}
\hspace{-0.3cm} dx \, [K_{\bar{\cal K}}]_{aa}(x,x;\beta)=
\sum_{a=1}^N\sum_{n=0}^\infty\frac{e^{-\beta\bar{v}_a^2}}{\sqrt{4\pi\beta}}
\int_{-\infty}^{\infty} \hspace{-0.3cm} dx \,
[a_n]_{aa}(x,x)\beta^n\\&=&\sum_{a=1}^N\sum_{n=0}^\infty\frac{e^{-\beta
\bar{v}_a^2 }}{\sqrt{4\pi\beta}} [a_n]_{aa}(\bar{\cal K}) \beta^n.
\end{eqnarray*}
The coefficients $[a_n]_{ab}(x,x')$ can be found by means of an
iterative procedure that starts from
$[a_0]_{ab}(x,x')=\delta_{ab}$ and becomes more and more involved
with larger and larger $N$. In Appendix A, this procedure is
explained and the explicit expressions for the lower orders are
shown.

The use of the power expansion of $h_{P\bar{{\cal K}}}[\beta]={\rm
Tr}e^{-\beta \,P \bar{{\cal K}}}$ in the formula for the quantum
kink mass is developed in three steps: \hspace{0.2cm}

1. First, we write the generalized zeta function of $\bar{\cal V}$
in the form:
\[
\zeta_{\bar{\cal V}}(s)=\frac{1}{\Gamma(s)} \frac{m_d L}{\sqrt{4
\pi}}\sum_{a=1}^N\int_0^1 d\beta \beta^{s-\frac{3}{2}}
e^{-\beta\bar{v}_a^2} +B_{\bar{\cal V}}(s)\quad ,
\]
with
\[
B_{\bar{\cal V}}(s)=\frac{m_dL}{\sqrt{4 \pi}}\sum_{a=1}^N
\frac{\Gamma[s-{\textstyle\frac{1}{2}},\bar{v}_a^2]}{\bar{v}_a^{2s-1}\Gamma[s]}
\hspace{0.4cm}, \hspace{0.4cm} \zeta_{\bar{\cal
V}}(s)=\frac{m_dL}{\sqrt{4\pi}}\sum_{a=1}^N
\frac{\gamma[s-{\textstyle\frac{1}{2}},\bar{v}_a^2]}{\bar{v}_a^{2s-1}\Gamma(s)}
+B_{\bar{\cal V}}(s);
\]
$\Gamma[s,\bar{v}^2]$ and $\gamma[s-\frac{1}{2},\bar{v}^2]$ being
respectively the upper and lower incomplete gamma functions, see
\cite{Abramowitz}. It follows that $\zeta_{\bar{\cal V}}(s)$ is a
meromorphic function of $s$ with poles at the poles of
$\gamma[s-{\textstyle\frac{1}{2}},\bar{v}_a^2]$, which occur when
$s-\frac{1}{2}$ is a negative integer or zero. $B_{\bar{\cal
V}}(s)$, however, is an entire function of $s$.

 2. Second, from the asymptotic expansion
of $h_{P\bar{\cal K}}[\beta]$ we estimate the generalized zeta
function of $\bar{\cal K}$:
\begin{eqnarray*}
\zeta_{P\bar{\cal K}}(s)&=&\frac{1}{\Gamma(s)} \left[-j\int_0^1
d\beta \beta^{s-1}+\frac{1}{\sqrt{4 \pi}} \sum_{a=1}^N\sum_{n<n_0}
[a_n]_{aa}(\bar{\cal K}) \int_0^1 d\beta \beta^{s+n-\frac{3}{2}}
e^{-\beta\bar{v}_a^2}+b_{n_0,\bar{\cal K}}(s) \right]
+B_{P\bar{\cal K}}(s)=\\ &=&-\frac{j}{s
\Gamma(s)}+\frac{1}{\Gamma(s) \sqrt{4 \pi}}\sum_{a=1}^N
\sum_{n<n_0} [a_n]_{aa}(\bar{\cal K})
\frac{\gamma[s+n-\frac{1}{2},\bar{v}_a^2]}{\bar{v}_a^{2(s+n-\frac{1}{2})}}+
\frac{1}{\Gamma(s)}b_{n_0,\bar{\cal K}}(s)+B_{P\bar{\cal K}}(s),
\end{eqnarray*}
where
\[
b_{n_0,\bar{\cal K}}(s)=\frac{1}{\sqrt{4\pi}}\sum_{a=1}^N
\sum_{n\geq n_0}^\infty [a_{n}]_{aa}(\bar{\cal K})
\frac{\gamma[s+n-\frac{1}{2},\bar{v}_a^2]}{\bar{v}_a^{2(s+n-\frac{1}{2})}}
\]
is holomorphic for ${\rm Re}\, s> -n_0+\frac{1}{2}$, whereas
\[
B_{P\bar{\cal K}}(s)=
\frac{1}{\Gamma(s)}\sum_{a=1}^N\int_1^\infty \,\
d\beta \,{\rm Tr} [e^{-\beta\,P{\cal K}}]_{aa} \beta^{s-1}
\]
is an entire function of $s$. The values of $s$ where $s+n-\frac{1}{2}$
is a negative integer or zero are the poles of $\zeta_{P\bar{\cal
K}}(s)$ because  the poles of
$\gamma[s+n-\frac{1}{2},\bar{v}_a^2]$ lie at these points in the complex $s$-plane.

Renormalization of the zero point energy requires the subtraction
of $\zeta_{\bar{\cal V}}(s)$ from $\zeta_{P \bar{\cal K}}(s)$. We
find:
\[
\zeta_{P\bar{\cal K}}(s)-\zeta_{\bar{\cal V}}(s) \approx
\frac{1}{\Gamma(s)} \left[
-\frac{j}{s}+\sum_{a=1}^N\sum_{n=1}^{n_0-1}
\frac{[a_n]_{aa}(\bar{\cal K})}{\sqrt{4 \pi}}
\frac{\gamma[s+n-\frac{1}{2},\bar{v}_a^2]}{\bar{v}_a^{2(s+n-\frac{1}{2})}}\right]\quad
,
\]
and the error in this approximation with respect to the exact
result to $\Delta_1\varepsilon^K$ is:
\[
{\rm error}_1=\frac{\hbar m_d}{2}
[-\textstyle\frac{1}{2\sqrt{\pi}} \,b_{n_0,\bar{\cal
K}}(-{\textstyle\frac{1}{2}})+B_{P\bar{\cal
K}}(-{\textstyle\frac{1}{2}})-B_{\bar{\cal
V}}(-{\textstyle\frac{1}{2}})] .
\]
Note that the subtraction of $\zeta_{\bar{\cal V}}(s)$ exactly
cancels the contribution of $a_0(\bar{\cal K})$ and hence the
divergence arising at $s=-\frac{1}{2}$, $n=0$.

\vspace{0.2cm}

3. Third, $\Delta_1\varepsilon^K$ now reads :
\begin{eqnarray*}
\Delta_1\varepsilon^K&\cong & \frac{\hbar
m_d}{\Gamma(-\frac{1}{2})} j+ \frac{\hbar}{2}
\lim_{s\rightarrow -\frac{1}{2}} \left(
\frac{\mu^2}{m_d^2}\right)^s \mu \,
\sum_{a=1}^N\frac{[a_1]_{aa}(\bar{\cal
K})}{\sqrt{4\pi}\Gamma(s)}\frac{
\gamma[s+{\textstyle\frac{1}{2}},\bar{v}_a^2]}{\bar{v}_a^{2s+1}}+\\
&&+\frac{\hbar m_d}{2}\sum_{a=1}^N \sum_{n=2}^{n_0-1}
\frac{[a_n]_{aa}(\bar{\cal K})}{\sqrt{4 \pi}\Gamma(-\frac{1}{2})}
\frac{\gamma[n-1,\bar{v}_a^2]}{\bar{v}_a^{2n-2}}\, .
\end{eqnarray*}
The surplus in energy due to the mass renormalization counter-term
is:
\begin{eqnarray*}
\Delta_2 \varepsilon^{\rm
K}&=&-\lim_{L\rightarrow\infty}\sum_{a=1}^N\frac{\hbar \,
[a_1]_{aa}(\bar{\cal K})}{2 L} \lim_{s\rightarrow -\frac{1}{2}}
\left( \frac{\mu}{m_d} \right)^{2 s+1}
\frac{\Gamma(s+1)}{\Gamma(s)}\zeta_{\bar{\cal V}_{aa}}(s+1) +o
(\hbar^2) \\ & \cong & -\frac{\hbar m_d}{2\sqrt{4\pi}}\,
\sum_{a=1}^N[a_1]_{aa}(\bar{\cal K}) \lim_{s\rightarrow
-\frac{1}{2}} \left( \frac{\mu}{m_d} \right)^{2 s+1}
\frac{\gamma[s+\frac{1}{2},\bar{v}_a^2]}{\bar{v}_a^{2s+1}\Gamma(s)}+o(\hbar^2)\,
,
\end{eqnarray*}
and the deviation from the exact result is :
\[
{\rm error}_2= \lim_{L\rightarrow\infty}\frac{\hbar}{4L}
\sum_{a=1}^N[a_1]_{aa}(\bar{\cal K}) B_{\bar{\cal
V}_{aa}}(\textstyle\frac{1}{2})\, .
\]
Therefore,
\begin{eqnarray*}
&&{\Bbb E}_R^{\rm K}=E[\psi_K]+\Delta M_K \cong
E[\psi_K]-\frac{\hbar m_d}{2\sqrt{\pi}} \left[
j+\frac{1}{4\sqrt{\pi}}\sum_{a=1}^N \sum_{n=2}^{n_0-1}
[a_n]_{aa}(\bar{\cal K})
\frac{\gamma[n-1,\bar{v}_a^2]}{\bar{v}_a^{2n-2}}\right]+\\ & &
+\frac{\hbar m_d}{2\sqrt{4 \pi}}\sum_{a=1}^N [a_1]_{aa}(\bar{\cal
K}) \lim_{s\rightarrow
-\frac{1}{2}}\left( \frac{\mu}{m_d}\right)^{2s+1} \,\left[
\frac{\gamma[s+\frac{1}{2},\bar{v}_a^2]}{\bar{v}_a^{2s+1}\Gamma(s)}-
 \frac{\gamma[s+\frac{1}{2},\bar{v}_a^2]}{\bar{v}_a^{2s+1}\Gamma(s)}\right]+o(\hbar^2)\quad
 .
\end{eqnarray*}
The contributions proportional to $[a_1]_{aa}(\bar{\cal K})$ of
the poles at $s=- \frac{1}{2}$ in $\Delta_1\varepsilon^K(s)$ and
$\Delta_2\varepsilon^K(s)$ exactly cancel.

\vspace{0.2cm}

We thus obtain the very compact formula:
\begin{equation}
\Delta M_K  \cong \hbar m_d \left[\Delta_0 +D_{n_0}\right] \left\{
\displaystyle\begin{array}{l} \Delta_0 =
-\displaystyle\frac{j}{2\sqrt{\pi}} \\ D_{n_0}=-\displaystyle
\sum_{a=1}^N
\sum_{n=2}^{n_0-1}\displaystyle\frac{[a_n]_{aa}(\bar{\cal
K})}{8\pi}\displaystyle\frac{\gamma[n-1,\bar{v}_a^2]}{\bar{v}_a^{2n-2}}
\end{array}
\right.  \quad .\label{eq:asy}
\end{equation}
In summ: there are only two contributions to semi-classical kink
masses obtained by means of the asymptotic method: 1) $\hbar m_d
\Delta_0$ is due to the subtraction of the zero modes; 2) $\hbar
m_d D_{n_0}$ comes from the partial sum of the asymptotic series
up to the $n_0$ order. We stress that the merit of the asymptotic
method lies in the fact that there is no need to solve the
spectral problem of ${\cal K}$: all the information is encoded in
the potential $V(x)$.

\section{The MSTB model: a deformed O(2) linear sigma system}

Let us consider now a $N=2$ model determined by the potential
energy density, \cite{Montonen}:
\begin{equation}
U[\vec{\psi}]=\frac{\lambda}{4}
(\vec{\psi}\vec{\psi}-\frac{m^2}{\lambda})^2+\frac{\alpha^2}{4}\psi_2^2\, .
\end{equation}
The system is a generalization of the $\lambda(\phi^4)_2$ model to
a $N=2$ scalar field, the $O(2)$-linear sigma model, although it
has been deformed by a quadratic term in $\psi_2$ in order to
avoid the Goldstone boson. The deformation parameter $\alpha$ has
dimensions of inverse length, $[\alpha]=L^{-1}$, and the choice of
non-dimensional variables in the so called MSTB model comes from
the choice of $c_d=\frac{\sqrt\lambda}{m}$,
$m_d=\frac{m}{\sqrt{2}}$ and $\alpha^2=m^2\sigma^2$ :
\[
\bar{U}(\phi_1,\phi_2)=\frac{1}{2}(\phi_1^2+\phi_2^2-1)^2+\frac{\sigma^2}{2}\phi_2^2\hspace{0.4cm}.
\]
$\sigma^2$ is the non-dimensional parameter that measures the
deformation and we shall also use the related parameter
$\bar{\sigma}^2=1-\sigma^2$. $\bar{U}(\vec{\phi)}$ is not
invariant under the $O(2)$-matrix group if $\sigma^2$ is not zero.
The \lq\lq internal" symmetry group is the ${\bf Z}_2\times {\bf
Z}_2$ group of discrete transformations:
\[
G_I=\left\{\vec{e}_a\rightarrow\pm\vec{e}_a\hspace{0.3cm},\hspace{0.3cm}
\vec{e}_a\rightarrow\pm (-1)^{\delta_{1a}}\vec{e}_a \right\}\quad
.
\]

The vacuum classical configurations are:
\[
\vec{\phi}_V(x,t)=\pm \vec{e}_1 \hspace{1cm},\hspace{1cm}
\vec{\psi}_V(y,y^0)=\pm\frac{m}{\sqrt{\lambda}}\vec{e}_1  .
\]
Therefore, if we denote by
\[
H_I^{(b)}=\left\{\vec{e}_a\rightarrow
\vec{e}_a\hspace{0.3cm},\hspace{0.3cm} \vec{e}_a\rightarrow
-(-1)^{\delta_{ba}}\vec{e}_a \right\}
\]
the ${\bf Z}_2$- subgroup that leaves $\vec{e}_b$ invariant, the
vacuum orbit and the vacuum moduli space are respectively: ${\cal
M}=\frac{G_I}{H_I^{(1)}}={\bf Z}_2$, $\hat{\cal M}=\frac{{\cal
M}}{G_I}={\rm point}$.

There are two kinds of topological kinks which are thus loop kinks
and candidates for being stable solitary waves of the system:
\begin{itemize}
\item Topological kinks with one non-null component:
\[
\vec{\phi}_{{\rm TK}1}(x,t)=({\rm
tanh}x)\vec{e}_1\hspace{1cm},\hspace{1cm}\vec{\psi}_{{\rm
TK}1}(y,y^0)=\frac{m}{\sqrt{\lambda}}({\rm
tanh}\frac{my}{\sqrt{2}})\vec{e}_1
\]
\item Topological kinks with two non-null components:
\[
\vec{\phi}_{{\rm TK}2}(x,t)=[({\rm tanh}\sigma x)\vec{e}_1\pm
\bar{\sigma} ({\rm sech}\sigma x )\vec{e}_2],
\]
\[\vec{\psi}_{{\rm TK}2}(y,y^0)=\frac{1}{\sqrt{\lambda}}[m({\rm
tanh}\frac{\alpha
y}{\sqrt{2}})\vec{e}_1\pm\sqrt{(m^2-\alpha^2)}({\rm
sech}\frac{\alpha y}{\sqrt{2}} )\vec{e}_2].
\]
Note that there are two two-component topological kinks , which
only exist if $0<\sigma^2<1$.
\end{itemize}

The kink and vacuum solutions have classical energies: ${\rm
E}[\vec {\psi}_{{\rm
TK}1}]=\frac{4}{3}\frac{m^3}{\sqrt{2}\lambda}$, ${\rm E}[\vec
{\psi}_{{\rm
TK}2}]=2\sigma(1-\frac{\sigma^2}{3})\frac{m^3}{\sqrt{2}\lambda}$
and ${\rm E}[\vec{\psi}_V]=0$. Thus, ${\rm E}[\vec {\psi}_{{\rm
TK}1}]>{\rm E}[\vec {\psi}_{{\rm TK}2}] $ and we shall see that
the ${\rm TK}1$ kink, as a quantum state, is unstable if
$\sigma^2<1$. The lower bound in energy in the topological sector
of the configuration space is attained, however, by the ${\rm
TK}2$ kink. This last statement is proved in Appendix B.

The Hessian operator for the vacuum solution is:
\[
{\cal V}= \left ( \begin{array}{cc} -\frac{d^2}{dy^2}+2m^2&0
\\ 0&
-\frac{d^2}{dy^2}+\frac{\alpha^2}{2}\end{array}
\right)=\frac{m^2}{2} \left (\begin{array}{cc}
-\frac{d^2}{dx^2}+4&0
\\ 0& -\frac{d^2}{dx^2}+\sigma^2\end{array}\right)=\frac{m^2}{2}\bar{\cal
V}\quad ,
\]
and hence the masses of the fundamental quanta are $
v_1^2=2m^2=\frac{m^2}{2}\bar{v}_1^2$ and
$v_2^2=\frac{\alpha^2}{2}=\frac{m^2}{2}\bar{v}_2^2$, in such a way
that the $H_I^{(2)}$ symmetry is spontaneously broken.

The Hessian operators for the topological kinks read:
\begin{itemize}
\item ${\rm TK}1$,
\[
\bar{\cal K}= \left( \begin{array}{cc}
-\frac{d^2}{dx^2}+4-\frac{6}{\cosh^2x}&0
\\ 0&
-\frac{d^2}{dx^2}+\sigma^2-\frac{2}{\cosh^2x}\end{array} \right)
\]

and ${\displaystyle{\cal K}=\frac{m^2}{2} \bar{\cal K}}$. Note that $\bar{{\cal K}}$ has a negative
eigenvalue if $\sigma^2<1$, as it should be for a unstable
solution.

\item ${\rm TK}2$,
\[
\bar{\cal H}= \left( \begin{array}{cc}
-\frac{d^2}{dx^2}+4-\frac{2(2+\sigma^2)}{\cosh^2\sigma
x}&4\bar{\sigma}\frac{{\rm tanh}\sigma x}{{\rm cosh}\sigma x}
\\ 4\bar{\sigma}\frac{{\rm tanh}\sigma x}{{\rm cosh}\sigma x}&
-\frac{d^2}{dx^2}+\sigma^2+\frac{2(2-3\sigma^2)}{\cosh^2\sigma
x}\end{array} \right)
\]
and ${\displaystyle{\cal H}=\frac{m^2}{2}\bar{\cal H}}$.
\end{itemize}
The corresponding generalized zeta functions satisfy
\[
\zeta_{\cal V}(s)=\left(\frac{2}{m^2}\right)^s\zeta_{\bar{\cal
V}}(s)\quad , \quad \zeta_{P{\cal
K}}(s)=\left(\frac{2}{m^2}\right)^s\zeta_{P{\bar{\cal K}}}(s)
\quad , \quad \zeta_{P{\cal
H}}(s)=\left(\frac{2}{m^2}\right)^s\zeta_{P{\bar{\cal H}}}(s)\quad
.
\]

\subsection{The quantum TK1 kink: exact computation of the semi-classical mass}
$\bullet$  Generalized zeta function of $\bar{\cal V}$:

Acting on the $L^2({\Bbb R})\otimes {\Bbb C}^2$ Hilbert space, we
have that
\[
{\rm Spec}\bar{\cal V}=\{k_1^2+4\}\cup \{k_2^2+\sigma^2\}={\rm
Spec}\bar{\cal V}_{11}\cup {\rm Spec}\bar{\cal V}_{22}  ,
\]
$k_1,k_2\in{\bf R}$, whereas the spectral density over a large
interval of length $L$ is :
\[
\rho_{\bar{\cal
V}}(k_1,k_2)=\left\{\begin{array}{ll}{\displaystyle\frac{mL}{\sqrt{2}\pi}}&{\rm
for }\ \ k_2^2\geq 4-\sigma^2 \\ {\displaystyle
\frac{mL}{2\sqrt{2}\pi}}&{\rm for }\ \ k_2^2<
4-\sigma^2\end{array}\right. \quad .
\]
From these data, the heat and generalized zeta functions can be
readily computed :
\[
{\rm Tr} e^{-\beta\bar{\cal V}}={\rm Tr} e^{-\beta\bar{\cal
V}_{11}}+{\rm Tr} e^{-\beta\bar{\cal
V}_{22}}=\frac{mL}{\sqrt{8\pi\beta}}\left(e^{-4\beta}+e^{-\sigma^2\beta}\right)ºquad
,
\]
\[
\zeta_{\bar{\cal V}}(s)=\zeta_{\bar{\cal
V}_{11}}(s)+\zeta_{\bar{\cal
V}_{22}}(s)=\frac{mL}{\sqrt{8\pi}}\left(\frac{1}{4^{s-\frac{1}{2}}}+
\frac{1}{(\sigma^2)^{s-\frac{1}{2}}} \right)\frac{\Gamma
(s-\frac{1}{2})}{\Gamma (s)} \quad .
\]

$\bullet$ Generalized zeta function of $\bar{\cal K}$

In the case of the one-component topological kink, computation of
the heat and generalized zeta functions is easy because $\bar{\cal
K}$ is diagonal. Moreover, $\bar{\cal K}_{11}$ and $\bar{\cal
K}_{22}$ are respectively the Hessian operators for the $\lambda
(\phi )^4$ kink and the sine-Gordon soliton (in the second case
shifted by $\sigma^2-1$). Therefore, we shall take advantage from
the work performed in \cite{AGM}. The spectrum of the Hessian
operator for the ${\rm TK}1$ kink is :
\[
{\rm Spec}\bar{\cal K}=\{0,\  3,\ \sigma^2-1\}\cup
\{k_1^2+4\}_{k_1\in{\bf R}}\cup \{k_2^2+\sigma^2\}_{k_2\in{\bf
R}}\quad ,
\]
and the spectral density and the phase shifts for the continuous
spectrum read :
\[
\rho_{\bar{\cal K}}(k_1,k_2)=\rho_{\bar{\cal
V}}(k_1,k_2)+\frac{1}{2\pi}\left(\frac{d\delta_1(k_1)}{dk_1}+
\frac{d\delta_2(k_2)}{dk_2}\right)
\]
\[
\delta_1 (k_1)=-2{\rm arctan}\frac{3k_1}{2-k_1^2}\hspace{1cm},\ \ \ \
\delta_2 (k_2)=2{\rm arctan}\frac{1}{k_2}
\]
respectively. Besides the continuous spectrum there are three
bound states with eigenvalues 0, 3 and $\sigma^2-1$. The
eigenfunction of 0 eigenvalue is the translational mode. There is
a second bound state in the $\vec{e}_1$ direction, but the third
bound state points along the $\vec{e}_2$ axis. Note that to
develop a non-zero $\phi_2$ component is energetically favorable
if $\sigma^2<1$ and this process costs energy when $\sigma^2>1$
and the ${\rm TK}1$ kink is stable. The second zero mode that
occurs at $\sigma^2=1$ is the signal of this phase transition.

The heat function
\begin{eqnarray*}
&&{\rm Tr}e^{-\beta P\bar{\cal K}}={\rm Tr}e^{-\beta P\bar{\cal
K}_{11}}+{\rm Tr}e^{-\beta\bar{\cal
K}_{22}}=\\&&=e^{-3\beta}+e^{-4\beta}\left[\int_{-\infty}^{\infty}\hspace{0.1cm}dk_1\left(\frac{mL}{\sqrt{8}\pi}+
\frac{1}{2\pi}\frac{d\delta(k_1)}{dk_1}\right)e^{-\beta
k_1^2}\right]+\\&&+e^{-(\sigma^2-1)\beta}+e^{-\sigma^2\beta}\left[\int_{-\infty}^{\infty}
\hspace{0.1cm}dk_2\left(\frac{mL}{\sqrt{8}\pi}+
\frac{1}{2\pi}\frac{d\delta(k_2)}{dk_2}\right)e^{-\beta
k_2^2}\right]
\end{eqnarray*}
is therefore equal to :
\[
{\rm Tr}e^{-\beta P\bar{\cal K}}={\rm Tr}e^{-\beta\bar{\cal
V}}+\left(e^{-3\beta}+e^{-(\sigma^2-1)\beta}\right){\rm
Erf}[\sqrt{\beta}]-{\rm Erfc}[2\sqrt{\beta}] \quad ,
\]
where ${\rm Erf}$ and ${\rm Erfc}$, \cite{Abramowitz}, are
respectively the error and complementary error functions. The
Mellin transform provides the corresponding zeta function:
\[
\zeta_{P\bar{\cal K}}(s)=\zeta_{\bar{\cal
V}}(s)+\frac{1}{\sqrt{\pi}}\left[\frac{2}{3^{s+\frac{1}{2}}}{}_2
F_1[{\textstyle\frac{1}{2}},s+\frac{1}{2},
{\textstyle\frac{3}{2}},-{\textstyle\frac{1}{3}}]-\frac{1}{4^s s}+
\frac{2}{(-\bar{\sigma}^2)^{s+\frac{1}{2}}}{}_2
F_1[{\textstyle\frac{1}{2}},s+\frac{1}{2},
{\textstyle\frac{3}{2}},{\textstyle\frac{1}{\bar{\sigma}^2}}]\right]\frac{\Gamma
(s+\frac{1}{2})}{\Gamma (s)} ,
\]
in terms of Gauss hypergeometric functions of the form
${}_2F_1[a,b,c,d]$, \cite{Abramowitz}.

Applying these results we obtain :
\begin{eqnarray*}
\Delta_1\varepsilon^K&=&\frac{\hbar}{2}\lim_{s\rightarrow
-\frac{1}{2}}\frac{1}{\sqrt{\pi}}\left(\frac{2\mu^2}{m^2}\right)^s\mu
 \\&\times&
\left[\frac{2}{3^{s+\frac{1}{2}}}{}_2F_1[{\textstyle\frac{1}{2}},s+\frac{1}{2},
{\textstyle\frac{3}{2}},-{\textstyle\frac{1}{3}}]-\frac{1}{4^s s}+
\frac{2}{(-\bar{\sigma}^2)^{s+\frac{1}{2}}}{}_2
F_1[{\textstyle\frac{1}{2}},s+\frac{1}{2},
{\textstyle\frac{3}{2}},{\textstyle\frac{1}{\bar{\sigma}^2}}]\right]\frac{\Gamma
(s+\frac{1}{2})}{\Gamma(s)},
\end{eqnarray*}
which is still a divergent quantity. The mass renormalization
counter-terms add another divergent quantity :
\begin{equation}
\Delta_2
\varepsilon^K=-\lim_{s\rightarrow-\frac{1}{2}}\lim_{L\rightarrow\infty}
\frac{2\hbar}{L}\left(\frac{2\mu^2}{m^2}\right)^{s+\frac{1}{2}}
\frac{\Gamma(s+1)}{\Gamma(s)}[3\zeta_{\bar{\cal
V}_{11}}(s+1)+\zeta_{\bar{\cal V}_{22}}(s+1)]\quad ,
\label{eq:massr1}
\end{equation}
and we obtain:
\begin{eqnarray*}
&&\Delta_1\varepsilon^K+\Delta_2\varepsilon^K=-\hbar\lim_{\varepsilon\rightarrow
0}
\left[\sqrt{\frac{1}{2\pi}}m\left(\frac{2\mu^2}{m^2}\right)^{\varepsilon}
\left(\frac{3}{4^{\varepsilon}}+\frac{1}{(\sigma^2)^{\varepsilon}}\right)
\frac{\Gamma (\varepsilon)}{\Gamma
(-\frac{1}{2}+\varepsilon)}\right]+\\&+&\frac{\hbar}{2}\lim_{\varepsilon\rightarrow
0}\left[\frac{m\sqrt{2}}{\sqrt{\pi}}\left(\frac{2\mu^2}{m^2}\right)^\varepsilon
\left(\frac{1}{3^{\varepsilon}}{}_2F_1[{\textstyle\frac{1}{2}},\varepsilon,
{\textstyle\frac{3}{2}},-{\textstyle\frac{1}{3}}]-
\frac{1}{4^\varepsilon(-\frac{1}{2}+\varepsilon)}+\frac{1}{(-\bar{\sigma}^2)^{\varepsilon}}{}_2
F_1[{\textstyle\frac{1}{2}},\varepsilon,
{\textstyle\frac{3}{2}},{\textstyle\frac{1}{\bar{\sigma}^2}}]\right)\right]\frac{\Gamma
(\varepsilon)}{\Gamma(-\frac{1}{2}+\varepsilon)} \, .
\end{eqnarray*}
It is appropriate to consider the contributions from ${\cal
K}_{11}$ and ${\cal K}_{22}$ separately before taking this limit:
\begin{itemize}
\item From ${\cal K}_{11}$ we obtain :
\begin{eqnarray*}
\Delta^{(1)}_1 \varepsilon^K &=& -\lim_{\varepsilon\to 0} \frac{3
\hbar m}{2\sqrt{2} \pi \epsilon} + \\ &+& \frac{\hbar m}{2
\sqrt{2} \pi} \left[ -4+3 \gamma_E+\log (3\cdot 2^4)-3 \log
\frac{2 \mu^2}{m^2}+3 \psi(\textstyle\frac{3}{2})-{}_2 F_1
'[\textstyle\frac{1}{2},0,\textstyle\frac{3}{2},-\textstyle\frac{1}{3}]
\right]+o(\epsilon)\, ,
\end{eqnarray*}
and
\[
\Delta^{(1)}_2 \varepsilon^K = \lim_{\varepsilon\to 0} \frac{3
\hbar m}{2 \sqrt{2} \pi \epsilon}+\frac{\hbar m}{2 \sqrt{2} \pi}
\left( -3 \gamma_E+3 \log \frac{2 \mu^2}{m^2}-6 \log 2-3
\psi(\textstyle\frac{3}{2}) \right)+o(\epsilon) \quad .
\]
Here, $\gamma_E$ is the Euler gamma constant; $\psi$ is the
Psi(Digamma) function, and ${}_2F_1'$ is the derivative of the
Gauss hypergeometric function with respect to the second argument,
\cite{Abramowitz} .
\item From ${\cal K}_{22}$ we obtain :
\begin{eqnarray*}
\Delta_1^{(2)} \varepsilon^K&=& -\lim_{\varepsilon\to 0}
\frac{\hbar m}{2\sqrt{2} \pi \epsilon} +
\\ &+& \frac{\hbar m}{2 \sqrt{2} \pi} \left[ \gamma_E-\log \frac{2
\mu^2}{m^2}+\log(\sigma^2-1)+\psi(\textstyle\frac{3}{2})-{}_2 F_1
'[\textstyle\frac{1}{2},0,\textstyle\frac{3}{2},\textstyle\frac{1}{1-\sigma^2}]
\right]+o(\epsilon)\, ,
\end{eqnarray*}
and
\[
\Delta^{(2)}_2 \varepsilon^K = \lim_{\varepsilon\to 0}\frac{\hbar
m}{2 \sqrt{2} \pi \epsilon}+\frac{\hbar m}{2 \sqrt{2} \pi} \left(
-\gamma_E+\log \frac{2 \mu^2}{m^2}-\log
\sigma^2-\psi(\textstyle\frac{3}{2}) \right)+o(\epsilon) \quad .
\]

\end{itemize}
Gathering all this together, we finally find:
\begin{eqnarray}
\Delta_1\varepsilon^K+\Delta_2\varepsilon^K &=&\Delta^{(1)}_1
\varepsilon^K+\Delta^{(1)}_2 \varepsilon^K +\Delta^{(2)}_1
\varepsilon^K+ \Delta^{(2)}_2 \varepsilon^K \nonumber \\ &=&
-\frac{\hbar
m}{2\sqrt{2}\pi}\left[4+\ln\frac{4}{3}+{}_2F_1'[{\textstyle\frac{1}{2}},0,
{\textstyle\frac{3}{2}},{\textstyle\frac{-1}{3}}]+\ln
(\frac{\sigma^2}{\sigma^2-1})+{}_2F_1'[{\textstyle\frac{1}{2}},0,
{\textstyle\frac{3}{2}},{\textstyle\frac{-1}{\sigma^2-1}}]\right]\nonumber\\&=&
-\frac{\hbar
m}{\pi\sqrt{2}}\left[(3-\frac{\pi}{\sqrt{12}})+(1-\sqrt{\sigma^2-1}\,
{\rm arcsin}\frac{1}{\sigma}) \right] \quad . \label{eq:hyp}
\end{eqnarray}

Note that $\Delta M_{{\rm
TK}1}(\sigma^2)=\Delta_1\varepsilon^K+\Delta_2\varepsilon^K$
acquires an imaginary part if $\sigma^2<1$ and in this latter
regime the quantum ${\rm TK}1$ kink becomes a resonance. Formula
(\ref{eq:hyp}) shows one interesting pattern in the one-loop
correction to the kink masses: First, the arguments of the
logarithms are always quotients of the eigenvalue where the
continuous spectrum starts by the energy of the bound eigen-state.
Second, the fourth argument of the derivatives of the Gauss
hypergeometric functions is always minus the inverse of the bound
-state eigenvalues.

\subsection{ The quantum topological kinks: asymptotic expansions}
The difficulty in computing the quantum correction to the mass of
the ${\rm TK}2$ kink lies in the fact that the $\bar{\cal H}$
Hessian operator at this configuration is a Schr\"{o}dinger
operator for which the $2\times 2$ matrix-valued potential $\bar{
V}(x)$ is non-diagonal. Therefore, the spectral problem of
$\bar{\cal H}$ is very difficult to solve and we are led to use
asymptotic methods to calculate the quantum ${\rm TK}2$ kink mass.

In the MSTB model there are two regimes:
\begin{enumerate}
\item  $\sigma^2<1$: The TK2 kink exists and is stable. In this
case we have that
\[
{\displaystyle\bar{
V}_{12}(x)=\bar{V}_{21}(x)=4\bar{\sigma}\frac{{\rm tanh}\sigma
x}{{\rm cosh}\sigma x}}
\]
is an odd function of $x$ such that $[a_1]_{12}(\bar{\cal
H})=[a_1]_{21}(\bar{\cal H})=0$. Bearing in mind that
$\bar{V}_{11}(x)=-\frac{2(2+\sigma^2)}{{\rm cosh}^2\sigma x}$ and
$\bar{V}_{22}(x)=-\frac{2(2-3\sigma^2)}{{\rm cosh}^2\sigma x}$, we
present the coefficients $[a_n]_{aa}(\bar{\cal H})$ up to $n_0=11$
with the help of Mathematica in the Tables 1 and 2 for several
values of $\sigma$:

\begin{table}[htbp]
\begin{center}
{\scriptsize
\begin{tabular}{|c|c|c|c|c|c|c|} \hline
&
\multicolumn{2}{|c|}{$\sigma=0.5$}&\multicolumn{2}{|c|}{$\sigma=0.7$}
& \multicolumn{2}{|c|}{ $\sigma=0.8$ } \\ \hline $n$
\rule[-0.2cm]{0cm}{0.6cm} & $[a_n]_{11}(\bar{\cal H})$ &
$[a_n]_{22}(\bar{\cal H})$ & $[a_n]_{11}(\bar{\cal H})$ &
$[a_n]_{22}(\bar{\cal H})$ & $[a_n]_{11}(\bar{\cal H})$ &
$[a_n]_{22}(\bar{\cal H})$
\\ & \rule{1.6cm}{0cm} & \rule{1.6cm}{0cm} & \rule{1.6cm}{0cm} &
\rule{1.6cm}{0cm} & \rule{1.6cm}{0cm} & \rule{1.6cm}{0cm}
\\[-0.4cm] \hline 1&17.9984 & -9.99909&14.2285&-3.02857&13.2000 &
-0.4000 \\ 2&34.9975 & 16.3313&27.5051&4.95579 &25.6320&2.42133\\
3&47.9636 &-16.8309 &38.5365&-3.88117&36.3858 & -1.46193\\
4&49.4219 & 13.2491 &40.9039&2.58248&39.2877 & 0.984184\\
5&40.6415 & -8.41779 & 34.5847&-1.43041&33.7529 & -0.53935\\
6&27.7911 & 4.4983 &24.2279&0.696572&23.9672&0.267282\\ 7&16.2632
& -2.08006 &14.4768&-0.30375&14.4819&-0.117869\\ 8&8.31759 &
-0.849979 &7.53962&0.119982&7.61204&0.0464771\\ 9&3.77771 &
-0.295903 &3.47957&-0.0425253&3.53989 &-0.0162638\\ 10&1.51992 &
0.0884623 &1.44002&0.0132872&1.47558 & 0.00497705\\ \hline
\end{tabular}}
\end{center}
\caption{\small {\it Heat function coefficients for several values
of the deformation parameter such that $\sigma < 0.9$ . }}
\end{table}
\begin{table}[htbp]
\begin{center}
{\scriptsize
\begin{tabular}{|c|c|c|c|c|} \hline
&
\multicolumn{2}{|c|}{$\sigma=0.9$}&\multicolumn{2}{|c|}{$\sigma=0.95$}
 \\ \hline $n$ \rule[-0.2cm]{0cm}{0.6cm}& $[a_n]_{11}(\bar{\cal H})$ & $[a_n]_{22}(\bar{\cal H})$ &
$[a_n]_{11}(\bar{\cal H})$ & $[a_n]_{22}(\bar{\cal H})$
\\ & \rule{1.6cm}{0cm} & \rule{1.6cm}{0cm} & \rule{1.6cm}{0cm} &
\rule{1.6cm}{0cm}
\\[-0.4cm] \hline 1&12.4889 & 1.91111&12.2211&2.97895 \\ 2&24.5218 & 1.67378&24.1951&1.952444\\
3&35.3368 &-0.241656 &35.16.16&0.310882\\ 4&38.8216 & 0.341514
&38.9553&0.230323\\ 5&33.8715 & -0.16798 & 34.2227&-0.053986\\
6&24.3610&0.0863412&24.7475&0.0372813\\
7&14.8742&-0.0370868&15.1758&-0.0145382\\
8&7.88516&0.0140749&8.07351&0.00547229\\ 9&3.69263
&-0.0046464&3.79191&-0.00173035\\
10&1.54847&0.001313&1.594006&0.0003535\\ \hline
\end{tabular}
\begin{tabular}{|c|c|c|} \hline
& \multicolumn{2}{|c|}{ $\sigma \geq 1$ }
\\ \hline $n$\rule[-0.2cm]{0cm}{0.6cm}  & $[a_n]_{11}(\bar{\cal K})$ & $[a_n]_{22}(\bar{\cal K})$
\\ & \rule{1.6cm}{0cm} & \rule{1.6cm}{0cm}
\\[-0.4cm] \hline 1&12.0000 &
4.0000 \\ 2&24.0000&2.66667\\ 3&35.2000 &1.066667\\ 4&39.3143
&0.304762\\ 5&34.7429 &0.0677249\\ 6&25.2306&0.0123136\\
7&15.5208&0.0018944\\ 8&8.27702&0.000252587\\
9&3.89498&2.97134\,$10^{-5}$\\ 10&1.63998&3.12591\,$10^{-6}$\\
\hline
\end{tabular}}
\end{center}
\caption{\small {\it Heat function coefficients for several values
such that $\sigma\geq 0.9$. If $\sigma\geq 1$ the coefficients
coming from $\bar{\cal K}$ are shown .}}
\end{table}

We also need to know that $j=1$: there is only one translational
zero mode in the spectrum of ${\cal H}$, and that $\bar{v}_1^2=4$
, $\bar{v}_2^2=\sigma^2$. Plugging this information in the
asymptotic formula (\ref{eq:asy}) we obtain the values shown in
Table 3 for $\Delta M_{{\rm TK}2}$.

\item $\sigma>1$: The only solitary wave is the ${\rm TK}1$ kink,
which is stable. To determine the quantum ${\rm TK}1$ kink mass by
using the asymptotic method, the general formulas must be applied
to $\bar{\cal K}$. Now, $\bar{V}_{12}(x)=\bar{V}_{21}(x)=0$ and
trivially $[a_1]_{12}(\bar{\cal K})=[a_n]_{21}(\bar{\cal K})=0$.
The $[a_n]_{aa}(\bar{\cal K})$ must be read from the formulas in
the Appendix applied to $\bar{V}_{11}(x)=-\frac{6}{{\rm
cosh}^2\sigma x}$ and $\bar{V}_{22}(x)=-\frac{2}{{\rm
cosh}^2\sigma x}$, and are shown in Table 2 up to $n_0=11$.
\begin{table}[htbp]
\begin{center}
\begin{tabular}{cc}
\begin{tabular}{|c|c|} \\[-0.6cm] \hline
$\sigma$ & $\Delta M_{{\rm TK}2}$ \\ \hline  $0.4$ & $-1.103270
\hbar m$
\\ $0.5$ & $-0.852622 \hbar m$ \\ $0.6$ & $-0.689001 \hbar m$
\\ $0.7$ & $-0.583835 \hbar m$ \\ $0.8$ & $-0.524363 \hbar m$ \\
$0.9$ & $-0.505708 \hbar m$ \\ $0.95$ & $-0.511638 \hbar m$
\\ \hline
\end{tabular} \hspace{1cm}
& \begin{tabular}{|c|c|} \\[-0.6cm] \hline $\sigma$ & $\Delta
M_{{\rm TK}1}$ \\ \hline $1.0$ & $-0.528311 \hbar m$ \\ $1.2$ &
$-0.518426\hbar m$ \\ $1.4$ & $-0.509645 \hbar m$
\\ $1.6$ & $-0.502291 \hbar m$ \\ $1.8$ & $-0.496369 \hbar m$ \\
$2.0$ & $-0.49172 \hbar m$ \\ $2.5$ & $-0.484183 \hbar m$ \\ $3.0$
& $-0.480101 \hbar m$ \\ $4.0$ & $-0.476181 \hbar m$  \\ \hline
\end{tabular}
\end{tabular}
\end{center}
\caption{\small {\it One-loop corrections to the mass of
topological kinks}}
\end{table}

Note that in this case the coefficients of the asymptotic
expansion of the generalized zeta function of $\bar{\cal K}$ are
independent of $\sigma$. Nevertheless, the quantum correction to
the ${\rm TK}1$ kink mass depends on $\sigma$ through the factors
$\frac{\gamma[n-1,\bar{v}_2^2]}{\bar{v}_2^{2n-2}}$. Again $j=1$:
there is only one translational zero mode in the spectrum of
${\cal K}$, and $\bar{v}_1^2=4$ , $\bar{v}_2^2=\sigma^2$.
Application of formula (\ref{eq:asy}) provides the values of
$\Delta M_{{\rm TK}1}$, also  shown in Table 3 for several values
of $\sigma$.

\end{enumerate}
We summarize the results obtained in this Section in the next
Figure, where $\Delta M_K$ is depicted as a function of $\sigma$.
It is understood that for $\sigma > 1$ we plot $\Delta M_{{\rm
TK}1}$ and for $0<\sigma <1$ $\Delta M_{{\rm TK}2}$ is
represented. The continuous line shows the exact value of $\Delta
M_{{\rm TK}1}$ as a function of $\sigma$, whereas the dots
correspond to the answer provided by the asymptotic method to
$\Delta M_{{\rm TK}1}$, (whites ), and $\Delta M_{{\rm TK}2}$,
(blacks ), for several values of $\sigma $.
\begin{figure}[htbp]
\centerline{\epsfig{file=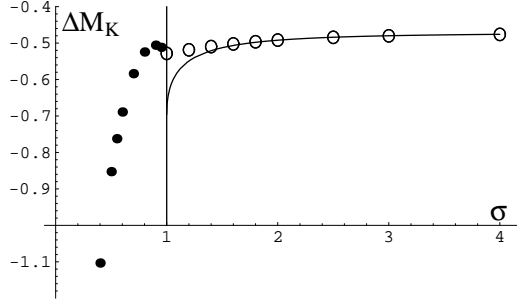,height=4cm}} \caption{\small
{\it Quantum correction to the mass in the MSTB model: TK2
($0<\sigma^2<1$, black dots) and TK1 ($\sigma^2\geq 1$, white
dots).}}
\end{figure}

We observe the following facts:
\begin{itemize}
\item In the $\sigma\geq 1.2$ range, the approximation established
by the asymptotic method is extremely good : the discrepancy with
the exact answer is compatible with zero.
\item In the $1<\sigma<1.2$ interval the error is larger and can
be estimated exactly in terms of $B_{\bar{\cal
K}^*}(-\frac{1}{2})$, $b_{n_0,\bar{\cal K}}(-\frac{1}{2})$,
$B_{\bar{\cal V}}(-\frac{1}{2})$ and $B_{\bar{\cal
V}_a}(\frac{1}{2})$. The closer the value of $\sigma$ to 1, the
smaller the first eigenvalue in the spectrum of $\bar{\cal K}^*$
and the larger $B_{\bar{\cal K}^*}(-\frac{1}{2})$. The asymptotic
method is better when fluctuations in the $\phi_2$ direction of
the ${\rm TK}1$ kink cost more energy.
\item The point $\sigma=1$, where the error is maximum, is
singular. There is a second zero eigenvalue which is the signal
for an instability-type phase transition on the ${\rm TK}1$ kink.
The co-existing unstable ${\rm TK}1$ and stable ${\rm TK}2$ kinks
for $\sigma<1$ coalesce to a single ${\rm TK}1$ kink at
$\sigma=1$, which becomes stable for $\sigma>1$.
\item In the $0<\sigma<1$ regime there is no way of estimating the
error in the approximation to $\Delta M_{{\rm TK}2}$ obtained by
means of asymptotic methods because the exact value is not known.
We expect, however, that the result will be better for smaller
values of $\sigma$ because of the same qualitative argument as
above: the cost in energy for climbing from the ${\rm TK}2$ to the
${\rm TK}1$ kink is larger with smaller $\sigma$.
\item The value $\sigma^2=4$ is a very special one. In this case
the system is ${\cal N}=2$ pre-supersymmetric in the sense that
there exists a super-potential :
\[
\vec{W}(\vec{\phi})=W^1(\phi_1,\phi_2)\vec{e}_1+W^2(\phi_1,\phi_2)\vec{e}_2
\quad ,
\]
where
\[
W^1(\phi_1,\phi_2)=\phi_1-\frac{\phi_1^3}{3}+\phi_1\phi_2^2 \quad
, \quad
W^2(\phi_1,\phi_2)=\phi_2-\phi_1^2\phi_2+\frac{\phi_2^3}{3}
\]
and
\[
\frac{1}{2}(\vec{\phi}\vec{\phi}-1)^2+2\phi_2^2=
\frac{1}{2}\frac{\partial W^1}{\partial\phi_1}\frac{\partial
W^1}{\partial\phi_1}+\frac{1}{2}\frac{\partial
W^1}{\partial\phi_2}\frac{\partial W^1}{\partial\phi_2}=
\frac{1}{2}\frac{\partial W^2}{\partial\phi_1}\frac{\partial
W^2}{\partial\phi_1}+\frac{1}{2}\frac{\partial
W^2}{\partial\phi_2}\frac{\partial W^2}{\partial\phi_2}\quad .
\]
This is no more than
 the super-potential of the holomorphic
Wess-Zumino model,
\[
\frac{\partial W^1}{\partial\phi_1}=\frac{\partial
W^2}{\partial\phi_2}\quad , \quad \frac{\partial
W^1}{\partial\phi_2}=-\frac{\partial W^2}{\partial\phi_1}\quad ,
\]
see \cite{AGM1}, and we have thus calculated the quantum
correction to the kink in this important system.
\end{itemize}
\section{Quantum kinks in a deformed O(N) linear sigma model}
Let us finally consider an $O(N)$, $N\geq 3$ model determined by
the potential energy density:
\begin{equation}
U[\vec{\psi}]=\frac{\lambda}{4}
(\vec{\psi}\vec{\psi}-\frac{m^2}{\lambda})^2+\sum_{l=2}^N\frac{\alpha_l^2}{4}\psi_l^2ºquad
,
\end{equation}
which generalizes the system discussed in Sect. 3. Without loss of
generality, the deformation parameters are chosen in such a way
that $\alpha_2<\alpha_3<\ldots<\alpha_N$. Passing to
non-dimensional variables, the potential is:
\[
\bar{U}(\vec{\phi})=\frac{1}{2}(\vec{\phi}\vec{\phi}-1)^2+
\sum_{l=2}^N\frac{\sigma_l^2}{2}\phi_l^2\hspace{0.4cm} ,
\]
with $\sigma_2<\sigma_3<\ldots<\sigma_N$. The \lq\lq internal"
symmetry group is $({\bf Z}_2)^N$; the vacuum classical
configurations are :
\[
\vec{\phi}_V(x,t)=\pm \vec{e}_1 \hspace{1cm},\hspace{1cm}
\vec{\psi}_V(y,y^0)=\pm\frac{m}{\sqrt{\lambda}}\vec{e}_1  ,
\]
and the vacuum orbit and the vacuum moduli space are
respectively: ${\cal M}=\frac{G_I}{H_I^{(1)}}={\bf Z}_2$,
$\hat{\cal M}=\frac{{\cal M}}{G_I}={\rm point}$.

There is a very  rich  manifold of kinks in this model, see
\cite{AGM4}. We shall only study the one-component and
two-component topological kinks, which are loop kinks and
candidate stable solitary waves in this system:
\begin{itemize}
\item Topological kinks and antikinks with one non-null component:
\[
\vec{\phi}_{{\rm TK}1}(x,t)=\pm({\rm
tanh}x)\vec{e}_1\hspace{1cm},\hspace{1cm}\vec{\psi}_{{\rm
TK}1}(y,y^0)=\pm\frac{m}{\sqrt{\lambda}}({\rm
tanh}\frac{my}{\sqrt{2}})\vec{e}_1 \quad .
\]
\item Topological kinks and anti-kinks with two non-null components.
For each $n$ from 2 to $N$ such that $\sigma_n<1$, we have :
\[
\vec{\phi}_{{\rm TK}2\sigma_n}(x,t)=\pm[({\rm tanh}\sigma_n
x)\vec{e}_1\pm \bar{\sigma}_n ({\rm sech}\sigma_n x )\vec{e}_n],
\]
\[
\vec{\psi}_{{\rm TK}2\sigma_n}(y,y^0)=\frac{\pm 1}{\sqrt{\lambda}}[m({\rm
tanh}\frac{\alpha_n
y}{\sqrt{2}})\vec{e}_1\pm\sqrt{(m^2-\alpha_n^2)}({\rm
sech}\frac{\alpha_n y}{\sqrt{2}} )\vec{e}_n]
\]
where $\bar{\sigma}_n=\sqrt{1-\sigma_n^2}$.
\end{itemize}

The kink and vacuum solutions have classical energies: ${\rm
E}[\vec {\psi}_{{\rm
TK}1}]=\frac{4}{3}\frac{m^3}{\sqrt{2}\lambda}$, ${\rm E}[\vec
{\psi}_{{\rm
TK}2\sigma_n}]=2\sigma_n(1-\frac{\sigma_n^2}{3})\frac{m^3}{\sqrt{2}\lambda}$,
${\rm E}[\vec{\psi}_V]=0$. Thus, ${\rm E}[\vec {\psi}_{{\rm
TK}1}]>{\rm E}[\vec {\psi}_{{\rm TK}2\sigma_N}]>\ldots>{\rm
E}[\vec {\psi}_{{\rm TK}2\sigma_2}] $. The lower bound in energy
in the topological sector of the configuration space is attained
by the ${\rm TK}2\sigma_2$ kink. If $\sigma_2>1$, only the ${\rm
TK}1$ kink exists and is stable. If $\sigma_j<1<\sigma_{j+1}$, the
kinks ${\rm TK}1$, ${\rm TK}2\sigma_2,\ldots$, ${\rm TK}2\sigma_j$
exist, but only the ${\rm TK}2\sigma_2$ is stable, see
\cite{AGM2}.

The Hessian operators in non-dimensional variables are as follows:
\begin{itemize}
\item Vacuum:
\begin{eqnarray*}
\bar{\cal V}&=&{\rm diag}(\bar{\cal V}_{ll})\\ \bar{\cal
V}_{11}&=& -\frac{d^2}{dx^2}+4 ,\ \ \ \ \ \bar{\cal V}_{ll}=
-\frac{d^2}{dx^2}+\sigma_l^2 ,\ l\geq 2 \quad .
\end{eqnarray*}
\item TK1:
\begin{eqnarray*}
\bar{\cal K}&=&{\rm diag}(\bar{\cal K}_{ll})\\ \bar{\cal
K}_{11}&=&-\frac{d^2}{dx^2}+4-\frac{6}{\cosh^2x},\ \ \ \ \
\bar{\cal K}_{ll}=-\frac{d^2}{dx^2}+\sigma_l^2 -\frac{2}{\cosh^2
x},\ \ \ l\geq 2 \quad .
\end{eqnarray*}
\item TK2$\sigma_n$:
\[
\bar{\cal H}^{(n)}=\left(\begin{array}{cc}\bar{\cal H}_{ND}^{(n)}&0\\0&\bar{\cal H}_D^{(n)}\end{array}\right)
\]
\[
\bar{\cal H}^{(n)}_{ND}=\left( \begin{array}{cc}
-\frac{d^2}{dx^2}+4-\frac{2(2+\sigma_n^2)}{\cosh^2\sigma_n
x}&4\bar{\sigma}_n\frac{{\rm tanh}\sigma_n x}{{\rm cosh}\sigma_n
x}
\\ 4\bar{\sigma}_n\frac{{\rm tanh}\sigma_n x}{{\rm cosh}\sigma_n x}&
-\frac{d^2}{dx^2}+\sigma_n^2+\frac{2(2-3\sigma_n^2)}{\cosh^2\sigma_n
x}\end{array} \right)
\]
\[
\bar{\cal H}^{(n)}_D={\rm diag}(-\frac{d^2}{dx^2}+\sigma_l^2-\frac{2\sigma_n^2}{\cosh^2\sigma_nx}),\ \ \ \ l=\hat{1},2,3,\ldots,\hat{n},\ldots,N ,
\]
where the hat over an index means that this index is absent and rows and columns have been rearranged in such a way that $\bar{\cal H}^{(n)}_{ND}$ acts on the small deformations of $\phi_1, \phi_n$ and $\bar{\cal H}^{(n)}_D$ on the remaining ones.
\end{itemize}
These operators are direct sums of the Hessian operators used in
Sect.3. We can therefore take advantage of the calculations
already made to give the quantum corrections to the kinks of the
present model.

\subsection{The quantum TK1 kink: exact computation of the semi-classical mass}
The relevant quantities are:
\begin{itemize}
\item  Generalized zeta function of $\bar{\cal V}$:
\[
\zeta_{\bar{\cal V}}(s)=\zeta_{\bar{\cal
V}_{11}}(s)+\sum_{l=2}^N\zeta_{\bar{\cal
V}_{ll}}(s)=\frac{mL}{\sqrt{8\pi}}\left(\frac{1}{4^{s-\frac{1}{2}}}+
\sum_{l=2}^N\frac{1}{(\sigma_l^2)^{s-\frac{1}{2}}}\right)\frac{\Gamma
(s-\frac{1}{2})}{\Gamma (s)}\quad .
\]
\item  Generalized zeta function of $P\bar{\cal K}$ :
\begin{eqnarray*}
\zeta_{P\bar{\cal K}}(s)&=&\zeta_{\bar{\cal
V}}(s)+\frac{1}{\sqrt{\pi}}\left[\frac{2}{3^{s+\frac{1}{2}}}{}_2
F_1[{\textstyle\frac{1}{2}},s+\frac{1}{2},
{\textstyle\frac{3}{2}},-{\textstyle\frac{1}{3}}]-\frac{1}{4^s
s}\right]\frac{\Gamma (s+\frac{1}{2})}{\Gamma (s)} \\&+&
\frac{1}{\sqrt{\pi}}\sum_{l=2}^N\left[\frac{2}{(-\bar{\sigma}_l^2)^{s+\frac{1}{2}}}{}_2
F_1[{\textstyle\frac{1}{2}},s+\frac{1}{2},
{\textstyle\frac{3}{2}},{\textstyle\frac{1}{\bar{\sigma}_l^2}}]\right]\frac{\Gamma
(s+\frac{1}{2})}{\Gamma (s)} .
\end{eqnarray*}
\end{itemize}
Applying these results, we obtain :
\begin{eqnarray*}
&&\Delta_1\varepsilon^K=\frac{\hbar}{2}\lim_{s\rightarrow
-\frac{1}{2}}\frac{1}{\sqrt{\pi}}\left(\frac{2\mu^2}{m^2}\right)^s\mu
\left[\frac{2}{3^{s+\frac{1}{2}}}{}_2F_1[{\textstyle\frac{1}{2}},s+\frac{1}{2},
{\textstyle\frac{3}{2}},-{\textstyle\frac{1}{3}}]-\frac{1}{4^s.s}\right]\frac{\Gamma
(s+\frac{1}{2})}{\Gamma(s)}
 \\
&+& \frac{\hbar}{2}\lim_{s\rightarrow
-\frac{1}{2}}\frac{1}{\sqrt{\pi}}\left(\frac{2\mu^2}{m^2}\right)^s\mu\sum_{l=2}^N\left[
\frac{2}{(-\bar{\sigma}_l^2)^{s+\frac{1}{2}}}{}_2
F_1[{\textstyle\frac{1}{2}},s+\frac{1}{2},
{\textstyle\frac{3}{2}},{\textstyle\frac{1}{\bar{\sigma}_l^2}}] \right] \frac{\Gamma
(s+\frac{1}{2})}{\Gamma(s)} .
\end{eqnarray*}
The mass renormalization counter-terms are :
\[
\Delta_2 \varepsilon^K=-\lim_{s\rightarrow -\frac{1}{2}}
\lim_{L\rightarrow\infty}
\frac{2\hbar}{L}\left(\frac{2\mu^2}{m^2}\right)^{s+\frac{1}{2}}
\frac{\Gamma(s+1)}{\Gamma(s)}[3\zeta_{\bar{\cal V}_{11}}(s+1)+\sum_{l=2}^N\zeta_{\bar{\cal
V}_{ll}}(s+1)] .
\]
Therefore,
\begin{eqnarray}
\Delta_1\varepsilon^K+\Delta_2\varepsilon^K&=&-\frac{\hbar
m}{2\sqrt{2}\pi}\left[4+\ln\frac{4}{3}+{}_2F_1'[{\textstyle\frac{1}{2}},0,
{\textstyle\frac{3}{2}},{\textstyle\frac{-1}{3}}]+\sum_{l=2}^N\left(\ln
(\frac{\sigma_l^2}{\sigma_l^2-1})+{}_2F_1'[{\textstyle\frac{1}{2}},0,
{\textstyle\frac{3}{2}},{\textstyle\frac{-1}{\sigma_l^2-1}}]\right)\right]\nonumber
\\&=&-\frac{\hbar
m}{\pi\sqrt{2}}\left[(3-\frac{\pi}{\sqrt{12}})+\sum_{l=2}^N(1-\sqrt{\sigma_l^2-1}\,
{\rm arcsin}\frac{1}{\sigma_l}) \right] . \label{eq:gen}
\end{eqnarray}
The ${\rm TK}1$ kink is a bona fide quantum state only if
$\sigma^2_l>1$, $\forall l \geq 2$.

\subsection{Quantum masses of two-component
 topological kinks: asymptotic expansion}
In the $\sigma_2^2 <1$ regime only the ${\rm TK}2\sigma_2$ kink is
stable. In this sub-section, therefore, we shall present the
calculation of the one-loop quantum correction to the ${\rm
TK}2\sigma_2$ kink mass. Since ${\rm Tr}e^{-\beta\bar{\cal
H}^{(2)}}={\rm Tr}e^{-\beta\bar{\cal H}_{ND}^{(2)}}+{\rm
Tr}e^{-\beta \bar{\cal H}_D^{(2)}}$, we encounter two old friends:
the heat functions arising respectively in connection with the
${\rm TK}2$ kink in the MSTB model and the soliton of the
sine-Gordon model. Using the information collected in previous
Sections, we find:
\begin{itemize}
\item Casimir energy :
\begin{eqnarray*}
\Delta_1\varepsilon^K &\cong&\frac{\hbar m}{2\sqrt{2}}
\lim_{s\rightarrow -\frac{1}{2}} \left( \frac{2 \mu^2}{m^2}
\right)^{s}\mu \left(-\frac{1}{s\Gamma (s)}
+\sum_{a=1}^2\sum_{n=1}^{n_0-1} \frac{a_n^{aa}(\bar{\cal
H}_{ND}^{(2)})}{\bar{v}_a^{2s+1}}
\frac{\gamma[s+n-\frac{1}{2},\bar{v}_a^2]}{\sqrt{4 \pi} \Gamma
(s)}\right)+\\&+&\frac{\hbar m}{2\sqrt{2}} \lim_{s\rightarrow
-\frac{1}{2}} \left( \frac{2 \mu^2}{m^2} \right)^{s}\mu
\sum_{l=3}^N\left(
\frac{2}{(\sigma_l^2-\sigma_2^2)^{s+\frac{1}{2}}}{}_2
F_1[{\textstyle\frac{1}{2}},s+\frac{1}{2},
{\textstyle\frac{3}{2}},{\textstyle\frac{-1}{\sigma_l^2-\sigma_2^2}}]\frac{\Gamma
(s+\frac{1}{2})}{\sqrt{\pi}\Gamma(s)}\right)\quad .
\end{eqnarray*}
\item Mass renormalization energy :
\begin{eqnarray*}
\Delta_2 \varepsilon^K &\cong& -\frac{\hbar
m}{\sigma_2\sqrt{2\pi}}\lim_{s\rightarrow-\frac{1}{2}}\left(
\frac{2 \mu^2}{m^2}
\right)^{s+\frac{1}{2}}\left(\frac{2+\sigma_2^2}{4^{s+\frac{1}{2}}}
\frac{\gamma [s+\frac{1}{2},4]}{\Gamma (s)}+
\frac{2-3\sigma_2^2}{\sigma_2^{2s+1}} \frac{\gamma
[s+\frac{1}{2},\sigma_2^2]}{\Gamma (s)}\right.\\&+&\left.\sigma_2^2 \sum_{l=3}^N \frac{\gamma
[s+\frac{1}{2},\sigma_l^2]}{\sigma_l^{2s+1}\Gamma (s)}\right) .
\end{eqnarray*}
\end{itemize}
We finally obtain the answer:
\begin{eqnarray*}
\Delta M_{{\rm TK}2\sigma_2}(\sigma_2,\sigma_l)&\cong&
-\frac{\hbar
m}{2\sqrt{2\pi}}\left(1+\frac{1}{4\sqrt{\pi}}\sum_{a=1}^2\sum_{n=1}^{n_0-1}
[a_n]_{aa}(\bar{\cal H}_{ND}^{(2)})\frac{\gamma
[n-1,\bar{v}_a^2]}{\bar{v}_a^{2n-2}}\right)-\\&-& \frac{\hbar
m}{2\sqrt{2\pi}}\sum_{l=3}^N\left(\log\frac{\sigma_l^2-\sigma_2^2}{\sigma_l^2)}+
{}_2F_1'[{\textstyle\frac{1}{2}},0,{\textstyle\frac{3}{2}},{\textstyle\frac{-1}
{\sigma_l^2-\sigma_2^2}}]\right) \quad ,
\end{eqnarray*}
or
\begin{eqnarray*}
\Delta M_{{\rm TK}2\sigma_2}(\sigma_2,\sigma_l)&\cong&
-\frac{\hbar
m}{2\sqrt{2\pi}}\left(1+\frac{1}{4\sqrt{\pi}}\sum_{a=1}^2\sum_{n=1}^{n_0-1}
[a_n]_{aa}(\bar{\cal H}_{ND}^{(2)})\frac{\gamma
[n-1,\bar{v}_a^2]}{\bar{v}_a^{2n-2}}\right)-\\&-& \frac{\hbar
m}{\sqrt{2}\pi}\sum_{l=3}^{N}\left(\sigma_2-\sqrt{(\sigma_l^2-\sigma_2^2)}{\rm
arcsin}\frac{\sigma_2}{\sigma_l} \right)\quad .
\end{eqnarray*}
In both formulas the contribution coming from $\bar{\cal
H}_{ND}^{(2)}$ can be read from the information on the quantum
correction to the ${\rm TK}2$ kink mass in the MSTB model,
collected in the Tables and the Figure of sub-Section \S.3.2 . A
similar formula would show us that $\Delta M_{{\rm TK}2\sigma_l}$,
$\forall l \geq 2$, receives an imaginary contribution any ${\rm
TK}2\sigma_l$ quantum kink state is therefore a resonance.

\section{Further comments}
The family of deformations that we have treated admits a ${\cal
N}=1$ super-symmetric extension. The super-potential
$\bar{W}(\phi)$ in the $N=2$ case is:
\[
\bar{W}(\phi)=\pm \sqrt{(\phi_1^2\pm\sigma
)^2+\phi_2^2}\,\,[\frac{1}{3}(\phi_1^2+\phi_2^2\mp\sigma\phi_1+\sigma^2)-1]\quad
.
\]
For the special value $\sigma=2$, this system also admits the
super-potential mentioned at the end of sub-Section \S 3.2 because
it becomes the Wess-Zumino model. If $N\geq 3$, it is very
difficult to write the super-potential in Cartesian coordinates in
the ${\Bbb R}^N$ internal space; nevertheless, passing to elliptic
coordinates one obtains easy expressions for the super-potential,
see \cite{AGM2}.

The super-symmetric extensions also include Majorana spinor
fields:
\[
\vec{\chi}(x^\mu)=\left(\begin{array}{c}\vec{\chi}_1(x^\mu)\\\vec{\chi}_2(x^\mu)\end{array}\right)\hspace{0.5cm},
\hspace{0.5cm}\vec{\chi}_\alpha^*=\vec{\chi}_\alpha \quad ,\quad
\alpha=1,2
\]
Choosing the Majorana representation $\gamma^0=\sigma^2,
\gamma^1=i\sigma^1, \gamma^5=\sigma^3$ of the Clifford algebra
$\{\gamma^\mu,\gamma^\nu\}=2g^{\mu\nu}$ and defining the Majorana
adjoint $\vec{\bar{\chi}}=\vec{\chi}^t \gamma^0$, the action of
the super-symmetric model is:
\begin{eqnarray*}
S&=&\frac{1}{2c_d^2}\int
dx^2\left\{\partial_\mu\vec{\phi}\partial^\mu\vec{\phi}+i\vec{\bar{\chi}}\gamma^\mu
\partial_\mu\vec{\chi}
-\vec{\nabla}W\vec{\nabla}W-\vec{\bar{\chi}}\vec{\vec{\Delta}}W\vec{\chi}\right\}\quad
,
\\ \vec{\nabla}\hat{W}(x^\mu)&=&\sum_{a=1}^2\hat{\frac{\partial
W}{\partial \phi^a}}(x^\mu )\vec{e}_a \quad ; \quad
\vec{\vec{\Delta}}\hat{W}=\vec{\nabla}\otimes\vec{\nabla}\hat{W}=\sum_{a=1}^2\sum_{b=1}^2
\vec{e}_a\otimes\vec{e}_b\frac{\partial^2\hat{W}}{\partial\phi^a\partial\phi^b}\,
.
\end{eqnarray*}
The ${\cal N}=1$ super-symmetry transformation is generated on the
space of classical configurations by the Hamiltonian spinor
function :
\[
Q=\int
dx\left\{\gamma^\mu\gamma^0\vec{\chi}\partial_\mu\vec{\phi}+i\gamma^0\vec{\chi}\vec{\nabla}
W \right\}\quad .
\]
The components of the Majorana spinorial charge Q close the
super-symmetry algebra :
\begin{equation}
\{Q_\alpha,Q_\beta\}=2(\gamma^\mu\gamma^0)_{\alpha\beta}P_\mu-2i\gamma^1_{\alpha\beta}T
 . \label{alg}
\end{equation}
Their (anti)-Poisson bracket is given in (\ref{alg}) in terms of
the momentum $P_\mu$ and the topological central charge $T=|\int d
W|=|\int \vec{\nabla}W d\vec{\phi}|$ .

The chiral projections $Q_{\pm}=\frac{1\pm\gamma^5}{2}Q$ and
$\chi_{\pm}=\frac{1\pm\gamma^5}{2}\chi$ provide a very special
combination of the super-symmetric charges:
\[
Q_++Q_-=\int
dx\left\{(\vec{\chi}_+-\vec{\chi}_-)\frac{d\vec{\phi}}{dx}-(\vec{\chi}_-+\vec{\chi}_+)
\vec{\nabla}W\right\}\quad .
\]
$Q_++Q_-$ is zero for the classical configurations that satisfy
$\frac{d\vec{\phi}}{dx}=\mp \vec{\nabla}W$ and
$\vec{\chi}_{\pm}=0$ which are thus classical BPS states. In
Appendix B it is proved that the stable TK2 kinks are such BPS
states and besides the small bosonic fluctuations one must take
into account the small fermionic fluctuations around the kink in
order to compute the quantum correction to the kink mass in the
extended super-symmetric system. The fermionic fluctuations around
the kink configuration lead to other solutions of the field
equations if the $N\times N$ matrix Dirac equation
\[
\left\{i\gamma^\mu
\partial_\mu+\vec{\vec{\Delta}}W(\vec{\phi}_{\rm K}
) \right\}\delta_{F}\vec{\chi}(x,t)=0
\]
is satisfied. We multiply this equation for the adjoint of the
Dirac operator :
\[
\left\{-i\gamma^\mu\partial_\mu+\vec{\vec{\Delta}}W(\vec{\phi}_{\rm
K})\right\}
\left\{i\gamma^\mu\partial_\mu+\vec{\vec{\Delta}}W(\vec{\phi}_{\rm
K})\right\}\delta_{F}\vec{\chi}(x,t)=0 \quad ,
\]
and, due to the time-independence of the kink background, look for
solutions of the form:
$\delta_{F}\vec{\chi}(x,t)=\vec{f}_F(x;\omega)e^{i\omega t}$. This
is tantamount to solving the spectral problem
\[
\left\{-\frac{d^2}{dx^2}+\vec{\vec{\Delta}}W(\vec{\phi}_{\rm
K})\vec{\vec{\Delta}}W(\vec{\phi}_{\rm K}) \mp
i\gamma^1\vec{\nabla}W(\vec{\phi}_{\rm
K})\vec{\nabla}\otimes\vec{\vec{\Delta}}W(\vec{\phi}_{\rm
K})\right\} \vec{f}_F(x;\omega)=\omega^2\vec{f}_F(x;\omega)\, .
\]
Projecting onto the eigen-spinors of $i\gamma^1$,
\[
\vec{f}_F^{(1)}(x;\omega)=\frac{1+i\gamma^1}{2}\vec{f}_F(x;\omega)=\frac{1}{2}
\left(\begin{array}{c}\vec{f}_F^+(x;\omega)-\vec{f}_F^-(x;\omega)\\-\vec{f}_F^+(x;\omega)+
\vec{f}_F^-(x;\omega)
\end{array}\right)
\]
we end with the spectral problem:
\[
\left\{-\frac{d^2}{dx^2}+\vec{\vec{\Delta}}W(\vec{\phi}_{\rm
K})\vec{\vec{\Delta}}W(\vec{\phi}_{\rm K})\mp
\vec{\nabla}W(\vec{\phi}_{\rm
K})\vec{\nabla}\otimes\vec{\vec{\Delta}}W(\vec{\phi}_{\rm
K})\right\} \vec{f}_F^{(1)}(x;\omega)={\cal
K}\vec{f}_F^{(1)}(x;\omega)=\omega^2\vec{f}_F^{(1)}(x;\omega)
\]
for the same Schrodinger operator that governs the bosonic
fluctuations.

Therefore, generalized zeta function methods can also be used in
super-symmetric models to compute the quantum corrections to the
mass of BPS kinks. Great care, however, is needed in choosing the
boundary conditions on the fermionic fluctuations without spoiling
super-symmetry. We look forward to extending this research in this
direction.

\appendix
\section{Appendix: the matrix heat kernel expansion} In this
Appendix we describe how to find the coefficients of the
asymptotic expansion of the heat kernel associated with $\bar{\cal
K}$ and show the explicit expressions for them up to third order;
for a more complete treatment see \cite{Avra}. Formula
(\ref{mateq}) in the text tells us that the matrix elements of
$A(x,x';\beta)$ satisfy the $N^2$ coupled PDE:
\[
\left\{\frac{\partial}{\partial\beta}+\frac{x-x'}{\beta}\frac{\partial}{\partial
x} -\frac{\partial^2}{\partial x^2}\right\} [A]_{ab}(x,x';\beta)
=\sum_{c=1}^N\bar{V}_{ac}(x) [A]_{cb}(x,x';\beta)+(\bar{v}_b^2
-\bar{v}_a^2)[A]_{ab}(x,x';\beta)\, ,
\]
with the initial condition: $A_{ab}(x,x';0)=\delta_{ab}$. Plugging
the power expansion of $A_{ab}(x,x';\beta)$ into this system of
equations we find the recurrence relations:
\begin{eqnarray*}
n[a_{n}]_{ab}(x,x')+(x-x')\frac{\partial[a_{n}]_{ab}}{\partial
x}(x,x')&=&\frac{\partial^2[a_{n-1}]_{ab}}{\partial
x^2}(x,x')+\sum_{c=1}^N\bar{V}_{ac}(x))[a_{n-1}]_{cb}(x,x')\\
&+&(\bar{v}_b^2-\bar{v}_a^2)[a_{n-1}]_{ab}(x,x')\quad .
\end{eqnarray*}
In order to take the $x'\rightarrow x$ limit properly, we
introduce the notation:
\[
^{(k)}[A_n]_{ab}(x)=\lim_{x'\rightarrow
x}\frac{\partial^k[a_n]_{ab}}{\partial x^k}(x,x')\quad .
\]
Then, the recurrence relations in the $x'=x$ limit become:
\[
[a_{n}]_{ab}(x,x)=\frac{1}{n}\left\{^{(2)}[A_{n-1}]_{ab}(x)+\sum_{c=1}^N\bar{V}_{ac}(x)
[a_{n-1}]_{cb}(x,x)+(\bar{v}_b^2-\bar{v}_a^2)[a_{n-1}]_{ab}(x,x) \right\}
\]
We also need the secondary recurrence relations among the
$^{(k)}[A_n]_{ab}(x)$ derived from $k$-times differentiation of the
primary recurrence relations above:
\[
^{(k)}[A_n]_{ab}(x)=\frac{1}{n+k}\left\{^{(k+2)}[A_{n-1}]_{ab}(x)+
\sum_{c=1}^N\sum_{j=0}^k\left(\begin{array}{c} k \\
j\end{array}\right)\frac{\partial^j \bar{V}_{ac}}{\partial x^j}
^{(k-j)}
[A_{n-1}]_{cb}(x)+(\bar{v}_b^2-\bar{v}_a^2)^{(k)}[A_{n-1}]_{ab}(x)\right\}\,
.
\]
Notice that $^{(k)}[A_0]_{ab}(x)=\lim_{x'\rightarrow
x}\frac{\partial^k[a_0]_{ab}}{\partial
x^k}=\delta^{k0}\delta_{ab}$. Thus, the $^{(k)}[A_n]_{ab}(x)$ and,
hence, the $[a_n]_{ab}(x,x)$ can be generated recursively . The
three first coefficients are :
\begin{eqnarray*}
\left[ a_1 \right]_{ab}(x)&=&\bar{V}_{ab}(x)\\ \left[ a_2
\right]_{ab}(x)&=&\frac{1}{6}\bar{V}_{ab}^{(2)}(x)+\frac{1}{2}\left[\bar{V}^2\right]_{ab}(x)
+\frac{1}{2}(\bar{v}_b^2-\bar{v}_a^2)\bar{V}_{ab}(x) \\
\left[a_3\right]_{ab}(x)&=&\frac{1}{60}\bar{V}_{ab}^{(4)}(x)+\frac{1}{12}\left[\bar{V}^{(2)}(x)\bar{V}(x)\right]_{ab}+\frac{1}{12}\left[\bar{V}(x)\bar{V}^{(2)}(x)\right]_{ab}+\frac{1}{12}\left[\bar{V}^{(1)}(x)\bar{V}^{(1)}(x)\right]_{ab}\\&+&\frac{1}{6}\left[\bar{V}^3\right]_{ab}(x)+\frac{1}{12}(\bar{v}_b^2-\bar{v}_a^2)\left\{\bar{V}^{(2)}_{ab}(x)+2\left[\bar{V}^2\right]_{ab}(x)\right\}\\&+&
\frac{1}{6}(\bar{v}_b^2-\bar{v}_a^2)^2\bar{V}_{ab}(x)+\frac{1}{6}\sum_{c=1}^N(\bar{v}_b^2
-\bar{v}_c^2)\bar{V}_{ac}(x)\bar{V}_{cb}(x)\quad .
\end{eqnarray*}

We mention that, as happens in the scalar case \cite{Perelomov},
the diagonal terms $[a_n]_{aa}(x,x)$ can be interpreted as the
densities giving the infinite conserved charges of a matrix
Korteweg-de Vries equation; namely :
\begin{equation}
\frac{\partial{\bar{V}}}{\partial t} +3[\bar{V}\frac{\partial{\bar{V}}}{\partial x} +\frac{\partial{\bar{V}}}{\partial x}\bar{V}]+\frac{\partial^3{\bar{V}}}{\partial x^3}=0 ,\label{kdv1}
\end{equation}
where now the matrix potential evolves in \lq\lq time" $t$, $\bar{V}=\bar{V}(x,t)$. The reason is that (\ref{kdv1}) can be written as a Lax equation
\[
L_t+[L,M]=0
\]
for the operators
\begin{eqnarray}
L&=&-\frac{\partial^2}{\partial x^2}-\bar{V}\label{kdv2}\\
M&=&4\frac{\partial^3}{\partial
x^3}-3\bar{V}\frac{\partial}{\partial x}
-3\frac{\partial}{\partial x}\bar{V}+B(t)\quad .
\end{eqnarray}
with $B(t)$ arbitrary. Therefore, standard arguments \cite{Drazin}
guarantee that the time evolution ruled by (\ref{kdv1}) produces
an uniparametric isospectral transformation of the Sch\"{o}dinger
operator (\ref{kdv2}). Because the integrals $[a_n]_{aa}$ are
determined by the spectrum of (\ref{kdv2}), their invariance
follows.

\section{Appendix: BPS and non-BPS kinks}
This Appendix is devoted to characterize respectively the TK1 and
TK2 kinks as non-BPS and BPS states in the $N=2$ case, see
Reference \cite{AGM2}. We have seen in Section \S 5 that the model
admits four super-potentials. If $\beta_a=0,1\, ,a=1,2$ , the four
super-potentials
\begin{equation}
W^{(\beta_1,\beta_2)}(\phi_1 ,\phi_2)=(-1)^{\beta_1}
\sqrt{(\phi_1+(-1)^{\beta_2}\sigma)^2+\phi_2}[\frac{1}{3}(\phi_1^2+\phi_2^2
-(-1)^{\beta_2}\sigma\phi_1+\sigma^2)-1]\quad ,  \label{eq:supt4}
\end{equation}
satisfy:
\[
\frac{1}{2}\sum_{a=1}^2\frac{\partial W^{(\beta_1
,\beta_2)}}{\partial\phi_a}\frac{\partial W^{(\beta_1
,\beta_2)}}{\partial\phi_a}=\frac{1}{2}(\phi_1^2+\phi_2^2-1)^2+\frac{\sigma^2}{2}\phi_2^2
\quad .
\]
The energy for static configurations reads:
\[
E=\frac{1}{2}\int dx\sum_{a=1}^2\left(\frac{d\phi_a}{dx}-
\frac{\partial W^{(\beta_1
,\beta_2)}}{\partial\phi_a}\right)\left(\frac{d\phi_a}{dx}-\frac{\partial
W^{(\beta_1 ,\beta_2)}}{\partial\phi_a}\right)+\int dx
\sum_{a=1}^2 \frac{d\phi_a}{dx}\frac{\partial W^{(\beta_1
,\beta_2)}}{\partial\phi_a} \quad .
\]
The BPS kinks are the solutions of the ODE first-order system :
\begin{eqnarray}
 \frac{d\phi_1}{dx}&=&\frac{\partial W^{(\beta_1
,\beta_2)}}{\partial\phi_1}=\frac{{\left( -1 \right) }^{\beta_1}
\left[ (-1+\phi_1^2)(\phi_1+(-1)^{\beta_2}\sigma
   )+\phi_1\phi_2^2
  \right]}{\,
     {\sqrt{{{\phi_2}^2} + {{\left( \phi_1 +
              {{\left( -1 \right) }^{\beta_2}}\,\sigma \right)
              }^2}}}}\label{eq:bode1}
  \\
\frac{d\phi_2}{dx}&=&\frac{\partial W^{(\beta_1
,\beta_2)}}{\partial\phi_2}={\frac{{{\left( -1 \right)
}^{\beta_1}}\,\phi_2\,
     [-\bar{\sigma}^2 +
     \,{{\phi_2}^2}+\phi_1(\phi_1+(-1)^{\beta_2}\sigma)
        ]}{
     {\sqrt{{{\phi_2}^2} + {{\left( \phi_1 +
              {{\left( -1 \right) }^{\beta_2}}\,\sigma \right)
              }^2}}}}}\quad \cdot
               \label{eq:bode2}
\end{eqnarray}
From the values of the super-potential at the vacuum points
\[
W^{(\beta_1,\beta_2)}(\pm 1,0)=(-1)^{\beta_1}(1\pm
(-1)^{\beta_2}\sigma)(-1+\frac{1}{3}(1\mp
(-1)^{\beta_2}\sigma+\sigma^2))
\]
we calculate the Bogomolny bound
\begin{equation}
E^{BPS}=|\int dW^{(\beta_1 ,\beta_2 )}|=|W^{(\beta_1 ,\beta_2
)}(1,0)-W^{(\beta_1 ,\beta_2
)}(-1,0)|=2\sigma\left(1-\frac{\sigma^2}{3}\right) \quad
,\label{eq:bbound}
\end{equation}
which is saturated by the solutions of
(\ref{eq:bode1},\ref{eq:bode2}), the BPS kinks.

In the derivation of (\ref{eq:bbound}) we have used Stokes's
theorem. The foci $(\phi^F_1=-(-1)^{\beta_2}\sigma,\phi_2^F=0)$ of
the ellipse $\phi_1^2+\frac{1}{\bar{\sigma}^2}\phi_2=1$ in the
${\Bbb R}^2$ \lq\lq internal" space are branching points of
$W^{(\beta_1 ,\beta_2)}$. Therefore, we are legitimated to use
Stokes's theorem - and the Bogomolny bound is reached- only if the
kink configuration does not cross any of the two foci above
mentioned. The TK2 kinks live on the semi-ellipses $\phi_2^{{\rm
TK}2}=\pm\bar{\sigma}\sqrt{1-(\phi_1^{{\rm TK}2})^2}$ and
everything is fine: they are BPS kinks. The TK1 solutions are more
involved. If $\phi_2=0$ (\ref{eq:bode1}) reduces to:
\[
\frac{d\phi_1}{dx}=-(-1)^{\beta_1}(1-\phi_1^2)\frac{\phi_1+(-1)^{\beta_2}\sigma}
{|\phi_1+(-1)^{\beta_2}\sigma|}=- (-1)^{\beta_1} (1-\phi_1^2)\,
\left( \Theta(\phi_1+\sigma (-1)^{\beta_2})-\Theta(-\phi_1-\sigma
(-1)^{\beta_2})\right) \quad ,
\]
where $\Theta (u)$ is the Heaviside step function. Thus,
$\phi_1^{{\rm TK}1}(x)={\rm tanh}x$ is not solution of the
first-order equations on the whole real line; if
$\phi_1+(-1)^{\beta_2}\sigma<0$, we must choose $\beta_1=0$, and
$\beta_1=1$ otherwise. On the half-line $x\in (-\infty,-{\rm
arctanh}[(-1)^{\beta_2}\sigma]]$ the TK1 solution is the flow line
of ${\rm grad}W^{(0,\beta_2)}$ but it becomes the flow line of
${\rm grad}W^{(1,\beta_2)}$ on $x\in [-{\rm
arctanh}[(-1)^{\beta_2}\sigma],\infty )$. One can easily check
that the TK1 kink is a proper solution of the second-order
equations with energy given by a piece-wise application of
Stokes's theorem:
\[
E^{{\rm
TK}1}=|W^{(0,\beta_2)}(-1,0)-W^{(0,\beta_2)}(-(-1)^{\beta_2}\sigma,0)|+
|W^{(1,\beta_2)}(-(-1)^{\beta_2}\sigma
,0)-W^{(1,\beta_2)}(1,0)|=\frac{4}{3}
\]
In a super-symmetric extension of this model the corresponding
state would be not annihilated by any combination of the
super-symmetry generators built from one of the
$W^{(\beta_1,\beta_2)}$ super-potentials and, therefore, the TK1
kink is a non-BPS state in this system if $\sigma^2<1$.

It is shown in Reference \cite{AGM2} that only the stable TK2
kinks are BPS kinks in the $N\geq 3$ models; the proof has been
performed using elliptic coordinates for the field variables.


\begin{thebibliography}{99}

\addcontentsline{toc}{section}{References}

\bibitem{Shifman} M. Shifman, A. Vainshtein and M. Voloshin, Phys. Rev. ${\bf D59}$ (1999) 45016.

\bibitem{Losev} A. Losev, M. Shifman and A. Vainshtein, \lq\lq Single State Supermultiplet in 1+1 Dimensions", hep-th/0011027.

\bibitem{Montonen} C. Montonen, Nucl. Phys. {\bf B112}  (1976) 349; S. Sarker, S.E. Trullinger, A.R. Bishop, Phys. Lett. {\bf 59A} (1976) 255.

\bibitem{Rajar}  R. Rajaraman, Phys. Rev. Lett. {\bf 42} (1979)
200  and \lq\lq Solitons and Instantons", North Holland,
Amsterdam, 1982; H. Ito, Phys. Lett. {\bf A112} (1985)119


\bibitem{AGM4} A. Alonso Izquierdo, M.A. Gonzalez Leon and J. Mateos Guilarte,
                   Nonlinearity. {\bf 13} (2000) 1137.

\bibitem{Graham} N. Graham and R. Jaffe, Nucl. Phys. ${\bf B544}$ (1999)
432 and Nucl. Phys. ${\bf B549}$ (1999) 516.

\bibitem{Bordag} M. Bordag, J. Phys. {\bf A28}  (1995) 755; E. Elizalde et al,
{\sl \lq\lq Zeta regularization techniques with applications"}, Singapore, World
Scientifique, 1994.


\bibitem{AGM}  A. Alonso Izquierdo, W. Garcia Fuertes, M.A. Gonzalez Leon and J. Mateos Guilarte,
               \lq\lq Generalized Zeta Functions and One-loop Corrections to Quantum Kink Masses",
               hep-th/0201084, to appear in Nuclear Physics B.

\bibitem{Bor} M. Bordag, A. Goldhaber, P. van Nieuwenhuizen, and
D. Vassilevich, \lq\lq Heat kernels and zeta-function regularization for
the mass of the SUSY kink", hep-th/0203066

\bibitem{Gilkey} P. Gilkey, {\sl \lq\lq Invariance theory, the heat equation
and the Atiyah-Singer index theorem"}, Publish or Perish,
Delaware, 1984.

\bibitem{Dashen} R. Dashen, B. Hasslacher and A. Neveu, Phys. Rev. ${\bf D10}$ (1974) 4130 and Phys. Rev. ${\bf D12}$ (1975) 3424.

\bibitem{Kor} L. D. Faddeev and V. E. Korepin, Phys. Rep. ${\bf 42C}$ (1978) 1-87.


\bibitem{Dunne} G.V. Dunne, Phys. Lett. {\bf 467}B (1999) 238.

\bibitem{Coleman} S. Coleman, {\sl \lq\lq Aspects of Symmetry"},
Cambridge University Press, 1985, Chapter 6: {\sl \lq\lq Classical
Lumps and their Quantum Descendants"}.

\bibitem{Avra} I.G. Avramidi and R. Schimming, J. Math. Phys. {\bf 36} (1995) 5042.

\bibitem{Abramowitz} M. Abramowitz and I. Stegun, {\sl \lq\lq Handbook of mathematical functions with formulas, graphs and mathematical tables"}, Dover Publications, Inc., New York, 1992.

\bibitem{AGM1} A. Alonso Izquierdo, M.A. Gonzalez Leon and J. Mateos Guilarte,
               Physics Letters {\bf B 480} (2000) 373.

\bibitem{AGM2} A. Alonso Izquierdo, M.A. Gonzalez Leon and J. Mateos Guilarte,
Nonlinearity {\bf 15} (2002)1097, math-ph/0204041.

\bibitem{Perelomov} A.M. Perelomov and Y. B. Zel'dovich, {\sl \lq\lq Quantum mechanics: selected topics"}, World Scientific, Singapore (1998).

\bibitem{Drazin} P. Drazin and R. Johnson, {\sl \lq\lq Solitons: an introduction"}, Cambridge University Press, Cambridge, 1996.







\end{thebibliography}
\end{document}